\documentclass{article}

\usepackage[utf8]{inputenc}
\usepackage[T1]{fontenc}
\usepackage[english]{babel}

\usepackage[numbers,sort&compress]{natbib}

\usepackage{amsmath,amssymb,bm}

\usepackage{graphicx}
\usepackage{color}
\usepackage{xspace}
\usepackage{comment}

\usepackage{lmodern}

\usepackage{amsthm}

\theoremstyle{definition}

\usepackage{xr}
\makeatletter
\newcommand*{\addFileDependency}[1]{%
  \typeout{(#1)}%
  \@addtofilelist{#1}%
  \IfFileExists{#1}{}{\typeout{No file #1.}}%
}
\makeatother

\newcommand*{\myexternaldocument}[1]{%
  \externaldocument{#1}%
  \addFileDependency{#1.tex}%
  \addFileDependency{#1.aux}%
}
\myexternaldocument{Supplementary}

\newcommand{\Gth}{G_\text{th}}
\newcommand{\omth}{\omega_\text{th}}
\newcommand{\omg}{\omega_\text{g}}
\newcommand{\omem}{\omega_\text{em}}
\newcommand{\omres}{\omega_\text{res}}
\newcommand{\Gammam}{\Gamma_{\!\text{m}}}
\newcommand{\Gammag}{\Gamma_{\!\text{g}}}
\newcommand\Atilde{\stackrel{\rule{.9ex}{0ex}\sim}{\smash{\bm{\mathcal{A}}}\rule{0pt}{1.4ex}}\hskip-.6ex}

\begin{document}
\title{Emission dynamics and spectrum of a nanoshell-based plasmonic nanolaser spaser}
\author{
  Ashod Aradian\thanks{Univ. Bordeaux, CNRS, CRPP, UMR 5031, F-33600 Pessac, France. \texttt{ashod.aradian@crpp.cnrs.fr}} \and
  Andres Cathey\thanks{Max Planck Institute for Plasma Physics, Boltzmannstr. 2, 85748 Garching, Germany. \texttt{andres.cathey@ipp.mpg.de}} \and
  Karen Caicedo\thanks{ISASI-CNR, Napoli, Italy. \texttt{gcaicedo@asig.com.ec}} \and
  Milena Mora\thanks{Universidad San Francisco de Quito, Ecuador. \texttt{smora@asig.com.ec}} \and
  Nicole Recalde\thanks{Universidad San Francisco de Quito, Ecuador. \texttt{nrecalde@alumni.usfq.edu.ec}} \and
  Melissa Infusino\thanks{Universidad San Francisco de Quito, Ecuador. \texttt{minfusino@usfq.edu.ec}} \and
  Alessandro Veltri\thanks{Universidad San Francisco de Quito, Ecuador. \texttt{aveltri@usfq.edu.ec}}
}
\date{}

\maketitle

\begin{abstract}
We study theoretically the emission and lasing properties of a single nanoshell spaser nanoparticle, or plasmonic nanolaser, made of an active core (gain material) and a plasmonic metal shell. Based on an analytical framework coupling together time-dependent equations for the gain and the metal, we calculate the lasing threshold with the help of an instability analysis. We characterize the regime under the threshold, where the nanoshell behaves as an optical amplifier when excited by an incident probe field. We then investigate in depth the non-linear lasing regime above the threshold, under autonomous conditions (free lasing without external drive), by computing the system's dynamics both in the transient state and in the final steady state. We show that at threshold, the lasing starts at one frequency only, usually one of the plasmon resonances of the nanoshell; then as the gain is further raised, the emission widens to other frequencies. This differs significantly from previous findings in the literature, which found only one emission wavelength above threshold. We proceed to calculate the complete (maximal) emission spectrum of the nanolaser as well as its emission linewidth, both of which are evidenced to be affected by unusually strong frequency shifts (pull-out) effects. We find that the nanolaser emission is highly asymmetrical spectrally and only occurs on one side (high-frequency) of the plasmon resonance.  Finally, we show that the spectral position of the emission line can be tuned across the whole visible range, by changing the geometrical aspect ratio of the nanoshell. 
\end{abstract}

\vspace{1em}

\begin{center}
\noindent\rule{\textwidth}{0.4pt}\\
\textbf{Supplementary Information is included at the end of this PDF (after the References).}\\
\noindent\rule{\textwidth}{0.4pt}
\end{center}

\section{Introduction}

With the advent of nanotechnology, one key aspect of research efforts in the past 30 years has been to generate, shape and manipulate light at subwavelength scales, much below the diffraction limit. An instrumental physical phenomenon towards this aim are surface plasmon polaritons, which are localized excitons where the electromagnetic field is coupled to electronic oscillations in a metal~\cite{Maier:2007}. If the metallic structure is at the nanoscale, then the associated fields are also generated at that same scale, and depending on cases, they may or may not be able to produce far-field radiation.

However, the use of metals at optical frequencies inevitably comes at the cost of significant Ohmic losses hindering the performance of plasmonic devices. One way to partially circumvent this fundamental issue is to use a gain material (active medium), placed at a distance close enough to the metallic structure: the gain medium can then transfer energy to the metal radiatively and/or non-radiatively. This strategy has proved to significantly improve properties and amplify the responses of device in various applications~\cite{Berini:2012,Hess:2012, DeLuca:2012, Meng:2011, Infusino:2014, Qian:2017, Polimeno:2020,ShiliangQu:2015-2}.

When the quantity of gain provided by the active medium is in excess to losses in the plasmonic system, one may enter a regime of nanolasing, i.e. the generation of coherent light at the nanoscale via the stimulated amplification of plasmons. Starting with the seminal concept of the spaser introduced in 2003~\cite{ber03}, quickly followed by the first experimental realizations of lasing spasers in 2009~\cite{Noginov:2009,Oulton:2009}, the field of nanolasers has since been highly active, as is testified by the flurry of reviews published over recent years~\cite{Ellis:2024,Ma:2021,Azzam:2020,Liang:2020,Stockman:2020,Wu:2019,Xu:2019,Ma:2019,Yang:2017,Deeb:2017,Wang:2017}.

Amongst the variety of geometries and schemes described in the literature, plasmonic nanolasers based on metallic nanoparticles (combining localized plasmons with gain within a nanoparticle) are especially attractive~\cite{Wang:2017}, due to the ease of mass fabrication of such structures with the help of bottom-up colloidal chemistry and self-assembly techniques~\cite{Baron:2021,Baron:2016}. One disadvantage of particle-based designs, however, is the difficulty to geometrically pack the necessary amount of gain to obtain the lasing~\cite{Wang:2017,Caicedo:2022}. In particular, there has been an ongoing debate about the true nature of the pioneering experiments by Noginov et al.~\cite{Noginov:2009} (lasing vs. random lasing, or some hybrid situation), and experimental realizations including metallic nanoparticles actually remain scarce~\cite{Wang:2017}.

In the wait for an improvement in the experimental situation, theoretical efforts have nonetheless been exploring lasing when nanoparticles are coupled to gain in various geometries: spheres, core-shells, multiple core-shells, ellipsoids…, using more or less refined analytical descriptions, or numerical simulation tools like the finite-element method~\cite{Lawandy:2004,Veltri:2012,Pustovit:2015,Pustovit:2016,Veltri:2016,Kewes:2017,Zhang:2012,Zhu:2014,Passarelli:2016,Wang:2020,Cuerda:2016,Szenes:2021, Vass:2024,Purohit:2024,Baranov:2013,Arnold:2015,Andrianov:2013,Khurgin:2012,Andrianov:2015,Parfenyev:2012,Parfenyev:2014,Andrianov:2011,Bordo:2017}. Most of these works were carried out assuming that the electromagnetic response for the materials involved could be faithfully accounted for using standard electrical permittivities; in particular for the gain medium, either a linear, Lorentzian permittivity or a non-linear saturated version were employed. When such an assumption is made from the start, however, it results in leaving out situations where more complex temporal dynamics may deploy. Since the latter is not uncommon in lasers, it is therefore necessary to rather make use of a fully time-dependent description to obtain a full understanding of the lasing regime. A few numerical studies have integrated all time and space-dependent effects using four-level population dynamics for gain carriers locally coupled to Maxwell equations~\cite{Cuerda:2016,Szenes:2021,Vass:2024,Purohit:2024}. Powerful as these are, full-wave simulations are not always transparent in terms of understanding the physical mechanisms at work, and a complementary model-based approach is undoubtedly useful.

In a past work~\cite{Veltri:2016}, we studied the situation of a metallic sphere immersed in an unbounded gain medium within the help of a time and space-dependent model. We proved the existence of a lasing threshold as the onset of an instability arising under zero driving field, starting first with the dipolar mode, and discussed how a mode cascade would subsequently occur due to a spatial hole-burning effect similar to laser physics, triggering many higher multipolar modes into emission. This multi-modal complication linked to the chosen geometry prevented us from studying the complete nanolaser’s dynamics above threshold.

\begin{figure}[h!]
\centering
 \includegraphics[width=0.7\columnwidth]{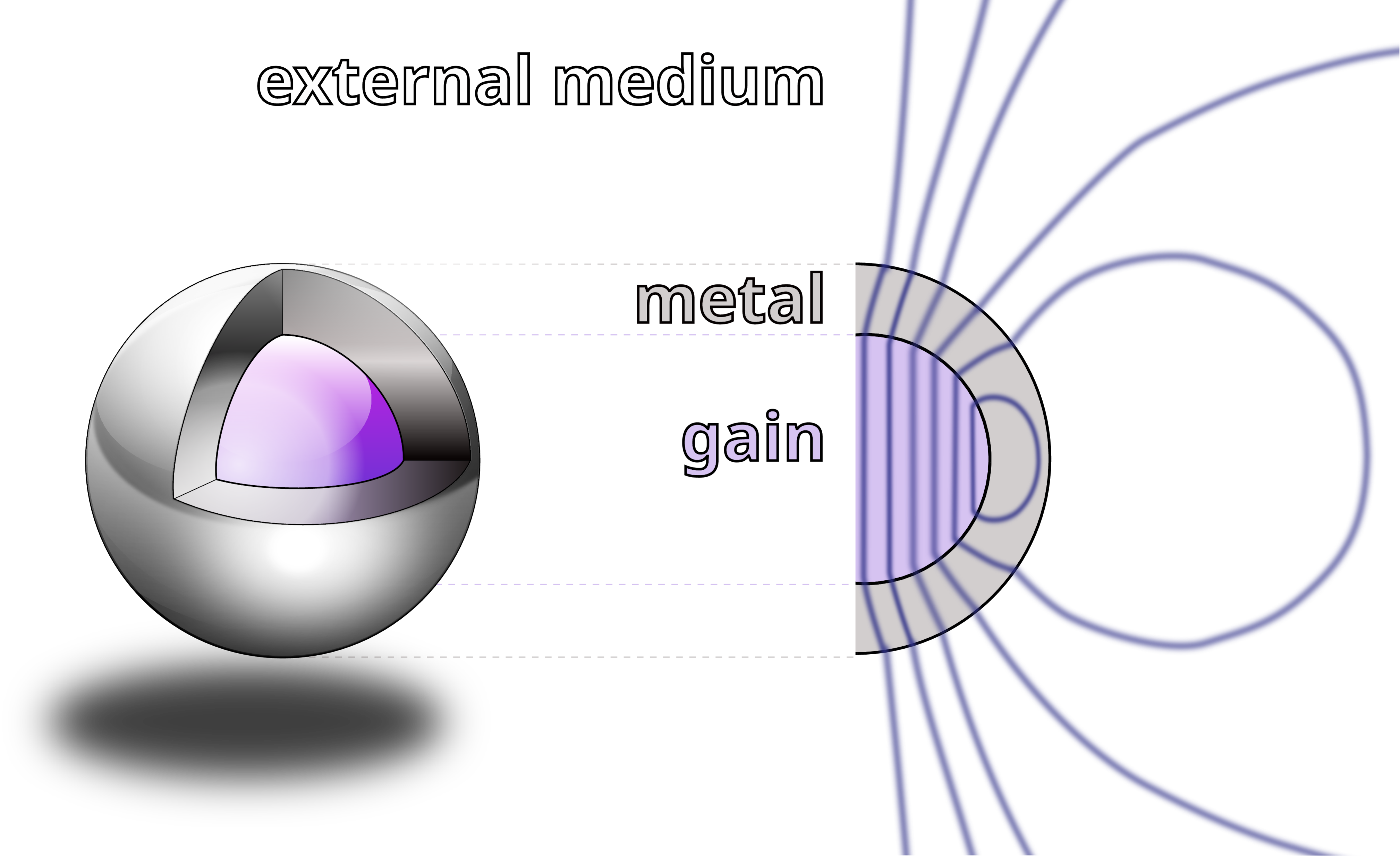}
\caption{A spherical nanoshell particle, with a gain medium filling the core and a metallic shell, placed into an external medium.}\label{fg:sp1}
\end{figure}

In the present article, we consider another geometry, closer to experiments, namely a single nanoshell where the gain medium is placed inside the core of the nanoparticle and is surrounded by a thin shell of metal (as shown in Fig.~\ref{fg:sp1}). In this case, as shall be explained, the dipolar is the only one that can emit in the lasing regime, making the analysis easier; based on the same model as earlier, we are then able to provide a full model-based characterization of the nonlinear lasing state for a nanoparticle, unveiling specific novel effects which, to the best of our knowledge, have remained unnoticed in the literature.

Finally, and before we start presenting the model, a word on vocabulary: in the following, we shall make no distinction between the intrinsically entangled notions of ``spasing’’ and ``lasing’’. In the specific context of nanoparticle-based plasmonic lasers, these notions are two sides of the same coin: the phenomenon of spasing focuses on the coherent excitation of plasmons in the nanoparticle, while lasing focuses on the emission of photons in the far-field associated with these plasmon oscillations~\cite{Ning:2021,Stockman:2020}. In this article, we shall mostly use the latter wording, as we will be primarily concerned with the emission of light by the nanoshell particle.

\section{Dynamical equations for materials}\label{sc:dce}
To ensure that our model is able to capture spatial and temporal variations of the fields, our first step is to formulate dynamical constitutive equations for the materials comprised in the nanoshell, i.e., for the metal and the gain medium. This formalism has been previously described in ref~\citenum{Veltri:2016}, and the interested reader is referred to the Supplementary Information of the present article for detailed derivations~\cite{SM}. We give below a summary of the important steps.

\subsection{Material equations}
The electron gas in the metal is classically modelled using the free-electron equation of motion~\cite{Veltri:2016,SM}, with a collision rate $\gamma$ and a plasma frequency $\omega_\text{p}$. The gain medium is described as a continuum composed of a background host material inside which gain elements (emitters), like dye molecules or quantum dots, are dispersed randomly. The population dynamics of electrons levels internal to the emitters is described with the help of an effective two-level approach, which is a phenomenological reduction capturing the essentials of more complete, multiple-level dynamics commonly used in laser physics~\cite{Chipouline:2012,Chubchev:2017,SM}. We label as levels 1 and 2, respectively, the lower and upper states of the resonant transition in the emitters, and call $\omg$ the angular frequency associated to this transition: $\omg=\Delta E_{21}/\hbar$, where $\Delta E_{21}$ is the energy gap between levels 1 and 2. The emitters are provided with energy by some external, optical pumping process, the details of which are left out: it is just assumed that the emitters are pumped with some tunable, effective pump rate $W$, at a frequency far from all phenomena of interest. It is also assumed that the metal shell is thin enough, and the frequency of the pump field high enough, that the pump wave penetrates inside the nanoparticle core. The population dynamics of this gain medium is then obtained with the help of the optical Bloch equations and the matrix density formalism~\cite{Chipouline:2012,Chubchev:2017,SM,Veltri:2016}.

Denoting $\mathbf r$ and $t$ as the spatial and time coordinates, the polarization ${\bf P}_\text{m}$ in the metal and ${\bf P}_\text{g}$ in the gain medium can be written as the sum of two contributions:
\begin{align}
 &{\bf P}_\text{g}(\mathbf r, t)=\epsilon_0\chi_\text{b}{\bf E}_\text{g}(\mathbf r, t)+{\bf\Pi}_\text{g}(\mathbf r, t),\label{ph}\\
 &{\bf P}_\text{m}(\mathbf r, t)=\epsilon_0\chi_\infty{\bf E}_\text{m}(\mathbf r, t)+{\bf\Pi}_\text{m}(\mathbf r, t).\label{pm}
\end{align}
Here, $\epsilon_0$ is the vacuum permittivity, $\chi_\text{b}$ is the background, passive linear susceptibility of the host medium in which the gain elements are dispersed, and $\chi_\infty$ is the background, passive susceptibility of the ion lattice in the metal.

The additional term ${\bf\Pi}_\text{g}$ is the polarization contribution from the dipole moments of the emitters within the host medium, and can be explicitly related to the transition dipole moment $\bm\mu$ between the two electronic levels as%
\begin{equation}\label{Pih}
{\bf \Pi}_\text{g}(\mathbf r, t)=\frac{n}{4\pi}\int_0^{\pi}\int_0^{2\pi}[\rho_{12}+\rho_{12}^*]{\bm\mu}\sin\theta d\theta d\varphi,
\end{equation}
where $n$ is the volumetric concentration of gain elements, $\rho_{12}$ is the element of the density matrix between levels 1 and 2, representing the gain element probability of transition, while $\theta$ and $\varphi$ are respectively the polar and the azimutal angle. The other additional term ${\bf\Pi}_\text{m}$ is the contribution to polarization due to the free electrons in the metal which is defined as
\begin{equation}\label{Pim}
 {\bf \Pi}_\text{m}(\mathbf r, t)=n_\text{e}e{\bf d},
\end{equation}
where $n_\text{e}$ and $e$ are respectively the electron density and the electron charge, and ${\bf d}$ is the displacement of the electron cloud with respect to the ionic lattice in the free-electron model. 

We now assume a harmonic form $e^{-i\omega t}$ (where $\omega$ is the angular frequency) for all time-dependent quantities. Due to the fundamental role of the gain medium in generating the lasing effect,  we shall assume that the operation frequency $\omega$ of the laser stays close to the frequency of the gain elements transition $\omg$. We will work within the frame of the rotating wave approximation~\cite{CohenTannoudji1998}, i.e., keeping track only of so-called "quasi-resonant" terms in equations, and considering slow temporal variations only with respect to the typical duration of optical cycles $\sim 1/\omega\sim 1/\omg$. Therefore, all fields and polarizations in the problem will be taken as slowly-varying complex quantities $\Atilde$, from which the corresponding, real-valued quantity ${\bm{\mathcal{A}}}$ in the physical world can be computed through the usual relation
\begin{equation}\label{eq:rea}
    \bm{\mathcal{A}}(\mathbf r, t)=\text{Re}[\Atilde(\mathbf r, t)e^{-i\omega t}].
\end{equation}
Henceforth, complex quantities $\Atilde(\mathbf r, t)$ will be considered only. However, for convenience, all tildas and the $(\mathbf r, t)$-dependence for fields will be implicitly assumed throughout the rest of this article and shall be dropped out of all equations.

%
%In order to carry on with our model we will also need to describe the field in the solvent in which the MNS is hosted, however being this material completely passive this would reduce to ${\bf P}_s=\epsilon_0\chi_s{\bf E}_s$ where $\chi_s$ is the susceptibility of the solvent.

Applying the optical Bloch equations for the gain part and the free-electron description for the metal, one obtains the following set of differential equations for the time evolution of the polarization inside the two materials in the nanoshell (see Supplementary Information~\cite{SM} for detailed steps): 
\begin{align}
      &\frac{d{\bf \Pi}_\text{g}}{dt}-\left[i(\omega-\omg)-\frac{1}{\tau_2}\right]{\bf \Pi}_\text{g}=-\frac{i\epsilon_0 G}{\tau_2}\frac{N}{\widetilde{N}}{\bf E}_\text{g}\label{eq:1ddn}\\
      &\frac{dN}{dt}+\frac{N-\widetilde{N}}{\tau_1}=-\frac{i}{2n\hbar}({\bf\Pi}_\text{g}\cdot{\bf E}^*_\text{g}-{\bf\Pi}_\text{g}^*\cdot{\bf E}_\text{g}) \label{eq:2ddn}\\
      &\frac{d{\bf \Pi}_\text{m}}{dt}-\frac{\omega(\omega+2i\gamma)}{2(\gamma-i\omega)}{\bf \Pi}_\text{m}=\frac{\epsilon_0\omega_\text{p}^2}{2(\gamma-i\omega)}{\bf E}_\text{m}.\label{eq:3ccn}
\end{align}
In the above, $N=\rho_{22}-\rho_{11}$ is the space and time-dependent population inversion of the emitters, defined as the difference of the diagonal terms of the density matrix. The time constant $\tau_2$ is associated with the phase relaxation processes of the emitters (collisions), while $\tau_1$ is the effective energy relaxation time, which results from the combined effect of spontaneous emission and pumping~\cite{SM,Veltri:2016}, $\widetilde{N}$ is the maximal, equilibrium value that the population inversion $N$ reaches as a balance between the applied pumping rate and spontaneous emission~\cite{Veltri:2016,SM}, in the absence of stimulated emission (i.e., if the saturation/depletion effects appearing in the r.h.s. of eq~\ref{eq:2ddn} are discarded).

\subsection{Gain level}
Finally, the quantity $G$ in eq~\ref{eq:1ddn} is a dimensionless parameter that will play a pivotal role in the following study~\cite{Caicedo:2022}:
\begin{equation}\label{epsmn}
 G =\frac{\tau_2\mu^2}{3\hbar\epsilon_0}n\widetilde{N}.
\end{equation}
We shall call $G$ the ``gain level'', indicative of the amount of total available power fed by the operator into to the system via the emitters and the external pumping. The value of $G$ is tunable; but once set to some desired value, it will be assumed to remain a constant parameter through the system's time evolution (constant pumping, or `CW' operation). As seen from its definition, the value of $G$ can be controlled as follows: $\tau_2$, $\mu$ depend on the choice of specific emitters (chemical dye, or otherwise) introduced in the gain medium; $n$ is the already defined volume density of emitters, and is limited by the quantity one can realistically pack into the nanoparticle's core; and finally, the quantity $\widetilde N$ is the most flexible one to change in practice, as it directly represents the external pumping rate $W$. More precisely, $\widetilde N=\widetilde N(W)$ is an increasing function of the pumping rate, with the extreme situations $\widetilde N = -1$ for zero pumping ($W=0$, all electrons in the lower level, purely absorbing medium) and $\widetilde N = 1$ for very intense pumping ($W \to \infty$, all electrons in the higher level, complete population inversion); see Supplementary Information~\cite{SM}, and ref~\citenum{Chipouline:2012} for details.

In ref~\citenum{Veltri:2016} about nanolasers made of plasmonic homogeneous spheres and in ref~\citenum{Caicedo:2022} about core-shell and nanoshell nanolasers, we discussed the existence of a threshold value $\Gth$, dependent on the physical properties of the gain elements and the system's geometry, such that when $G$ exceeds $\Gth$, the systems would transit into a lasing/spasing regime. In the case of the nanoshell geometry, the threshold gain was defined as the point where the classical formula for the quasi-static polarizability of the particle becomes singular. We shall demonstrate that this intuitive definition still holds true in the following, more elaborate model.  

The system of equations~\ref{eq:1ddn}--\ref{eq:3ccn} is applicable to a variety of situations that demand dynamical constitutive equations for both the metal and the gain medium. Its solutions encompass, in principle, all possible transient states both above and below the emission threshold, as well as steady states.

\subsection{Steady-state permittivities}
Before moving to the case of nanoshells, let us quickly review such steady-state solutions obtained in the case of infinite media subject to uniform electric fields. From equation~\ref{eq:3ccn}, one can retrieve the standard Lorentz-Drude formula for the permittivity of the metal (see Supplementary Information~\cite{SM}):
\begin{equation}\label{eq:dru}
 \epsilon_\text{m}=\epsilon_\infty-\frac{\epsilon_0 {\omega_\text{p}}^2}{\omega(\omega+2i\gamma)},
\end{equation}
 with $\epsilon_\infty=\epsilon_0 ( 1+\chi_\infty )$. 

For the gain medium modeled by equation~\ref{eq:1ddn} along with equation~\ref{eq:2ddn}, one finds the steady-state permittivity~\cite{SM}:
\begin{equation}\label{eq:slor}
 \epsilon_\text{g}=\epsilon_\text{b}+\frac{[2(\omega-\omg)-i\Delta]\epsilon_0 G \Delta}{4(\omega-\omg)^2+\Delta^2\left[1+\dfrac{\left|\mathbf{E}_\text{g}\right|^2}{{E_\text{sat}}^2}\right]},
\end{equation}
where we defined $\epsilon_\text{b}=\epsilon_0 ( 1+\chi_\text{b})$, $\Delta=2/\tau_2$, and
\begin{equation}\label{eq:esatt}
E_\text{sat}=\frac{\hbar}{\mu}\sqrt{\frac{3}{\tau_1\tau_2}}.   
\end{equation}
Equation~\ref{eq:slor} stands as a non-linear permittivity for the gain medium, dependent on the modulus of the electric field $\left|\mathbf{E}_\text{g}\right|$ inside it. This is classically known as a ``gain saturation'' effect due to the depletion term ${\bf\Pi}_\text{g}\cdot{\bf E}^*_\text{g}-{\bf\Pi}_\text{g}^*\cdot{\bf E}_\text{g}$ in eq~\ref{eq:2ddn}: when the field in the gain medium becomes large enough, the upper state of the resonant transition of the emitters becomes depleted, causing $N$ to decrease and saturate at some value lower than the maximum value $\widetilde N$ allowed by the pump. The typical magnitude of the field where this becomes significant is $E_\text{sat}$. In the ``small-signal'' regime where fields keep small enough with $\left|\mathbf{E}_\text{g}\right|^2 \ll {E_\text{sat}}^2$, the depletion term in eq~\ref{eq:2ddn} is negligible, and $N$ quickly converges to its maximum, unsaturated value $\widetilde N$; so that the above permittivity becomes a linear one:
\begin{equation}\label{eq:lor}
 \epsilon_\text{g}= \epsilon_\text{b}+\frac{\epsilon_0 G \Delta}{2(\omega-\omg)+i\Delta},
\end{equation}
which is no other than the widely used Lorentzian curve used for unsaturated gain media, centered at $\omega=\omg$, with an emission linewidth $\Delta$.

In the context of nanolasers, however, we will show that these steady-state permittivities for metal and gain have to be used with care above the threshold of lasing, especially in situations of free lasing (i.e., without any external drive) where the nanolaser is free to choose its frequency of emission.  
\section{Nanoshell model}
\subsection{Nanoshell description}
We now proceed to specify the geometry of the nanoshell shown in Figure~\ref{fg:sp1}, of external radius $r_\text{ext}=a$, internal radius $r_\text{int}=\rho a$ and aspect ratio $\rho=r_\text{int}/r_\text{ext}$. The core is filled with the gain medium described in the previous section (eq~\ref{eq:1ddn}--\ref{eq:2ddn}), and is surrounded by a shell made with the metal described by eq~\ref{eq:3ccn}. The whole nanoparticle is bathing in an external medium (i.e., a solvent) which is assumed to be a passive dielectric with a real, positive permittivity $\epsilon_\text{e}$.

We place ourselves in the quasi-static limit, where the nanoparticle's size is much smaller than the impinging wavelength. The exciting probe field can be approximated as spatially uniform and written as $\mathbf{E}_0e^{-i\omega t}$, where $\mathbf{E}_0$ does not depend on spatial position. Within the same approximation, we can introduce the time and space-dependent potentials $\phi_{g,m,e}$ and $\psi_{g,m}$, respectively located in the gain core ($g$), metal shell ($m$) and external medium ($e$), from which the fields and polarizations are derived as ${\bf E}_{g,m,e}=-\nabla\phi_{g,m,e}$ and ${\bf \Pi}_{g,m}=-\nabla\psi_{g,m}$. They must satisfy the Laplace equations:
\begin{align}
    &\nabla^2\phi_{g,m,e}=0\label{eq:laphi}\\
    &\nabla^2\psi_{g,m}=0\label{eq:lapsi}
\end{align}

While using the quasi-static approximation is perfectly admissible with small, passive nanoparticles, it is a much more delicate affair in the context of active, lasing nanoparticles. Indeed, for the fields to derive from potentials, they should be curl-free (irrotational), i.e. $\nabla\times{\bf E}_{g,m,e}=\nabla\times{\bf \Pi}_{g,m}=0$. But as can be seen from eq~\ref{eq:1ddn} where the righ-hand side contains an $N\cdot{\bf E}_\text{g}$ term, the spatial pattern of the inversion population $N$ can constitute a source of rotationality for the fields. This is discussed in detail in a previous work~\cite{Recalde:2023}, where it was shown that if the inversion population stays spatially uniform in the gain medium, i.e. $\nabla N={\bf 0}$, all fields remain irrotational in time (provided the size of the particle is small enough).

In all situations where the fields stay within the small-signal regime discussed above ($\left|\mathbf{E}_\text{g}\right|^2 \ll {E_\text{sat}}^2$), since $N(\mathbf r, t)=\widetilde N$ always, the latter condition $\nabla N={\bf 0}$ will be satisfied. This will be applicable to all geometrical arrangements of nanolasers below their lasing threshold, since their response is proportional to the intensity of the exciting probe field~\cite{Veltri:2016,Caicedo:2022}: provided the latter is not exceedingly intense, which we will assume further down, all fields remain small. (For cases where the probe field is very intense and the system falls out of the small-signal regime even below the lasing threshold, the reader is referred to ref~\citenum{Cerdan:2023}.)

On the contrary, if the system is placed above the lasing threshold, the fields grow considerably, and the uniformity of $N$ is not guaranteed due to spatial hole burning effects. In ref~\citenum{Veltri:2016}, we studied in depth how spatial burning occurs in the case of a nanolaser made of a homogeneous plasmonic sphere placed in an unbounded gain medium: irrespective of how small the particle is, in the lasing regime, $N$ inevitably becomes non-uniform and breaks the irrotational hypothesis; and, the quasi-static, dipolar mode of the particle in fact always excites a cascade of higher (non quasi-static) multipolar modes. Intuitively, this is because in that geometry, when the dipolar mode is initially active within the sphere, it produces a dipole field $\mathbf{E}_\text{g}$ (which is non-uniform by nature) in the external gain medium. Therefore, the r.h.s. of eq~\ref{eq:2ddn} is non-uniform, which in turn brings on a non-uniform evolution of $N$: the inversion population is then ``burnt'' according to the spatial pattern cut out by the laser field. The same spatial hole-burning scenario holds, for the same reasons, in a core-shell nanolaser, with the plasmonic metal inside the core and the gain medium in the shell.

Depending on their geometry, the presence of spatial hole burning even in small nanolasers is an important fact that is most often overlooked in the literature: many works incorrectly study core-shell nanoparticles in the lasing regime, taking only the dipolar mode into account due to size considerations; thus missing out on the physics of the multipolar mode cascade that will inevitably take place.

However, in the nanoshell geometry which we will exclusively consider in the present study, the situation is distincly different, because the gain medium is now \emph{inside} the core: when the dipole (quasi-static) mode of the nanoparticle is first activated, it creates a \emph{uniform} field $\textbf{E}_\text{g}$ inside the core, hence the depletion in the r.h.s. of eq~\ref{eq:2ddn} is uniform as well, which entails that $N$ will then keep uniform throughout its time evolution. This brings a substantial simplification, as irrotationality is preserved and a potential-based approach is here legitimate to describe the whole dynamics of the system, both below and above the lasing threshold. In a nanoshell geometry, the quasi-static dipolar mode remains alone, without any higher modes being excited~\cite{NonuniformNote}.

We use spherical coordinates centered on the nanoparticle, aligning the $z$-axis along the direction of the probe field (i.e.,  $\mathbf{E}_0=E_0\hat{\bf z}$), and assume azimuthal symmetry around $z$. Then, equations~\ref{eq:laphi} and~\ref{eq:lapsi} produce solutions that can be expressed as a superposition of Legendre polynomials. Taking into account that the potentials should be regular at $r=0$ and that for $r\gg 1$, the electric field has to reconnect to the probe field $\mathbf{E}_0$, the following expressions for~$\phi_{g,m,e}$ and~$\psi_{g,m}$ are obtained:
\begin{align}
\phi_\text{g}(r,\theta,t)&=p_0r\cos\theta,\label{ph1l1}\\
\phi_\text{m}(r,\theta,t)&=p_1r\cos\theta+a^3\rho^3p_2\frac{\cos\theta}{r^2},\label{ph2l1}\\
\phi_\text{e}(r,\theta,t)&=-E_0r\cos\theta+a^3p_3\frac{\cos\theta }{r^2},\label{ph3l1}\\\psi_\text{g}(r,\theta,t)&= q_0r\cos\theta,\label{ps1}\\
\psi_\text{m}(r,\theta,t)&=q_1r\cos\theta+a^3\rho^3q_2\frac{\cos\theta}{r^2}. \label{ps2}
\end{align}
Equations~\ref{ph1l1} and \ref{ps1} state that the field and polarization inside the core (gain medium) are uniform; equations~\ref{ph3l1} states that the external field is the sum of the probe field and the single-mode, dipolar field generated by the nanoparticle; equations~\ref{ph2l1} and \ref{ps2} state that the field and polarization in the metallic shell, are the sum of a constant field and a dipolar field. The coefficients $p_0$, $p_1$, $p_2$, and $p_3$ are the amplitudes of the constant and dipolar modes of the electrical fields in the various domains of the system, while $q_0$, $q_1$ and $q_2$ are the corresponding mode amplitudes for the polarizations. It is possible to link the values of the $q_i$ and $p_i$ variables through the imposition of the appropriate boundary conditions at the various interfaces of the nanoshell (see ``Methods'' section). 

In particular, we note that $p_0$ is the amplitude of the uniform field inside the gain-medium core: $p_0=E_\text{g}$; while $p_3$ is the amplitude of the dipolar field scattered by the nanoparticle in the external medium. It is proportional to the the total dipole moment $\mathcal{P}$ of the nanoparticle through 
\begin{equation}
    \mathcal{P}=4 \pi \epsilon_\text{e} a^3p_3.
    \label{eq:dipmoment}
\end{equation}

\subsection{Geometry matrix and governing equations}
To obtain the governing set of equations for the nanoshell's dynamics, we next introduce the above dipolar description of eq~\ref{ph1l1}--\ref{ps2} into the previously described material equations~\ref{eq:1ddn}--\ref{eq:3ccn}. With the help of intermediate steps described in the ``Methods'' section, we find that this governing set can be written in the following matrix form:
\begin{align}
 &\frac{d{\bf q}}{dt}={\bf A}(N)\cdot{\bf q}+{\bf b},\label{eq:qmastereq}\\
 &\frac{dN}{dt}+\frac{N-\widetilde{N}}{\tau_1}=\frac{1}{n\hbar}\text{Im}\left\{q_0p_0^*\right\}.\label{eq:Nmastereq}
\end{align}
This set of equations fully describes the time-dependent electrodynamical behavior of the nanoshell, in a self-contained fashion. The vector ${\bf q}$ collects the electromagnetic mode components:
\begin{equation}
    {\bf q}(t)=\bigl[q_0, q_1, q_2\bigr]^\mathrm{T},
\end{equation} 
and the complete physical definition of the system is contained within the matrix $\bf{A}$ and vector $\bf{b}$ (see ``Methods'' section for complete expressions). Solving Equation~\ref{eq:qmastereq}, one obtains the value of all polarizations components $q_i(t)$, and then all related field components $p_i(t)$, inside and outside the nanoshell, as a function of time. Solving the (coupled) equation~\ref{eq:Nmastereq}, one obtains the simultaneous evolution of the population inversion $N(t)$ in the gain region. 

It is useful to describe the physical contents of matrix ${\bf A}$, which is central to our model. We shall call ${\bf A}$ the ``geometry matrix'', as its components encode all the information about the nanoshell geometry of the system (see ``Methods''). As emphasized by the notation ${\bf A}(N)$, $\bf A$ also explicitly depends on the population inversion $N=N({\bf q},t)$, which in general is time-dependent, and, most importantly, depends non-linearly on the electromagnetic modes $q_i$. Therefore, the system~\ref{eq:qmastereq}--\ref{eq:Nmastereq} is a non-linear one in ${\bf q}$ in the most general case. Furthermore, the ${\bf A} (N)$ matrix features the level of gain $G$ brought to the system via pumping, as defined in eq~\ref{epsmn}, whose value will be critical to determine the various regimes of response of the nanoshell. Finally, ${\bf A} (N)$ depends on the frequency $\omega$.

The vector ${\bf b}(N,E_0)$ depends on $N$ as well, but moreover specifically carries the information on the excitation by the probe field $E_0$; in the absence of a probe field ($E_0=0$), one has $\bf{b}=\bf{0}$.

In the next sections, we will study the dynamical behaviour of the nanoshell laser, as dictated by eq~\ref{eq:qmastereq}--\ref{eq:Nmastereq}, both below and above the lasing threshold. We start off by evidencing the existence of such a threshold in our formalism.
\section{Lasing threshold}
\label{sc:thresh}
Classically, we define the lasing threshold as the minimal quantity of gain $G=\Gth$ to be provided to the system, in order to observe the rise of a self-oscillation of the nanoshell~\cite{Veltri:2016,Baranov:2013}, i.e., the rise of non-zero fields inside and outside the particle \emph{in the absence of an exciting probe field}. We thus take $E_0=0$ and $\bf{b}=\bf{0}$.

Right at the onset of the self-oscillation, fields are small and therefore the ``small-signal'' approximation is valid: one can neglect the r.h.s. of eq~\ref{eq:Nmastereq}. After a transient of duration $\sim\tau_1$, $N$ will reach the stationary value $N=\widetilde N$, which is independent of all variables $q_i$. Then, the geometry matrix becomes a constant matrix ${\bf A}(\widetilde N)$, meaning that eq~\ref{eq:qmastereq} is now a linear differential equation with generic solutions
\begin{equation}
    {\bf q}(t) = \sum_{n=1}^{n=3} {\bf \hat q}_n e^{\kappa_n t},
    \label{eq:diffsol}
\end{equation}
where $\kappa_n$ are the eigenvalues of ${\bf A}(\widetilde N)$ and ${\bf \hat q}_n$ are associated eigenvectors.

Above the threshold, the solution should be exponentially growing (i.e., at least one eigenvalue has a positive real part) as a result of the self-oscillation instability. Below the threshold, the solution should be on the contrary exponentially decaying (i.e., all eigenvalues have a strictly negative real part), as no field should emerge in the absence of external excitation (no self-oscillation). We refer the reader to ref~\citenum{Veltri:2016} for a similar analysis carried out in the case of a metal sphere immersed in a gain medium.

The lasing threshold is thus characterized by the tipping point where one null eigenvalue appears in the spectrum of ${\bf A}$. The lasing condition therefore simply writes:
\begin{equation}
    \det[{\bf A} (\widetilde N)] = 0.
    \label{eq:lasingcond}
\end{equation}
As this condition implies that both the real and imaginary part of $\det[{\bf A} (\widetilde N)]$ be cancelled simultaneously, it determines both the threshold gain value, $G=\Gth$, and the lasing frequency of the nanoshell at the threshold, $\omega=\omth$. 
After some cumbersome calculations (see~\cite{SM} for details), condition~\ref{eq:lasingcond} can be rewritten under the simple form:
\begin{equation}
    (\epsilon_\text{g}+2\epsilon_\text{m})(\epsilon_\text{m}+2\epsilon_\text{e})+2\rho^3(\epsilon_\text{g}-\epsilon_\text{m})(\epsilon_\text{m}-\epsilon_\text{e})=0,
    \label{eq:denominatorcond}
\end{equation}
which needs to be solved to find the couple $(\omth, \Gth)$.

We note that this condition is the same that we proposed in ref~\citenum{Caicedo:2022}, where a more intuitive argument was followed. It is useful to briefly remind of this argument, which proceeded from considering the classical formula for the quasi-static polarizability $\alpha$ for a nanoshell particle~\cite{bohren98}:
\begin{equation}
    \frac{\alpha}{4\pi a^3} = \frac{ \left( \epsilon_\text{g} + 2 \epsilon_\text{m} \right) \left( \epsilon_\text{m} - \epsilon_\text{e} \right) + \rho^3 \left( \epsilon_\text{g} - \epsilon_\text{m} \right) \left( \epsilon_\text{e} + 2 \epsilon_\text{m} \right) }{ \left( \epsilon_\text{g} + 2 \epsilon_\text{m} \right) \left( 2 \epsilon_\text{e} + \epsilon_\text{m} \right) + 2 \rho^3 \left( \epsilon_\text{g} - \epsilon_\text{m} \right) \left( \epsilon_\text{m} - \epsilon_\text{e} \right) }.
    \label{eq:polarizability}
\end{equation}
This polarizability allows to calculate the value of the nanoshell's total dipolar moment $\mathcal{P}$ when it is excited by a probe field $E_0$, as
\begin{equation}
    \mathcal{P} = \epsilon_\text{e} \alpha E_0.
    \label{eq:dipolalpha}
\end{equation}
The self-oscillation of the nanolaser is then defined as the situation when this dipolar moment remains finite ($\mathcal{P} \neq 0$) even if the probe field is made to vanish ($E_0 \to 0$). This is possible only if the polarizability $\alpha$ becomes singular at the lasing threshold, or in other words, if its denominator cancels out, which is exactly the same as condition~\ref{eq:denominatorcond}.

On a rigorous standpoint, the use of the classical expression~\ref{eq:polarizability} for polarizability remains unsubstantiated at this stage, but it will be justified fully in the next section ``Below threshold''.

It furthermore appears from the above argument on polarizability that the lasing frequency at threshold $\omth$ is in fact no other than one of the natural plasmon resonance frequencies $\omres$ of the nanoshell, which are usually also found by cancelling the denominator of $\alpha$, i.e.,
\begin{equation}
    \omth=\omres.
    \label{eq:omegares}
\end{equation}

Two important remarks are in order at this point. Firstly, these natural plasmon resonance frequencies are those of the nanoparticle \emph{in the presence of gain} inside the core, with the gain level set at $G=\Gth$. Therefore, the actual values found for $\omres$, and hence $\omth$, will depend on the level of gain and on the positioning of the centerline frequency $\omg$ of the gain emission spectrum, through the presence of $\epsilon_\text{g}(\omega)$ in the denominator of~\ref{eq:polarizability}. The optimal situation allowing the lowest lasing threshold is obtained when the gain provides its peak value to one of the resonances, i.e. it has been chosen to be centered exactly on the same frequency. In this case, comparing to eq~\ref{eq:omegares}, one has an additional equality:
\begin{equation}
    \omg = \omres = \omth \quad \text{(optimal gain)}.
    \label{eq:optimalgain}
\end{equation}
In the rest of this paper, we will always assume such optimal gain positioning. In this case, it can be shown that $\omres$ and $\omth$ are independent of the gain level, and are the same as the resonant frequencies of the nanoshell with zero gain (found by solving eq~\ref{eq:denominatorcond} with $\epsilon_\text{g}=\epsilon_\text{b}$). Note, however, that the resonant frequencies still depend on the nanoshell's aspect ratio $\rho$, so that the gain centerline should be adjusted each time $\rho$ is modified, to remain optimal. Effects linked to gain detuning (non-optimal positioning) with respect to the nanoshell resonances are left for future work.

Secondly, nanoshells exhibit two plasmonic resonances, one symmetric (lower frequency) and one anti-symmetric (higher frequency)~\cite{Maier:2007}. In principle, both resonances can be brought to lasing. In this work, we arbitrarily choose to focus on provoking the lasing of the symmetric resonance only, by centering the gain spectrum on the lowest frequency solution of eq~\ref{eq:denominatorcond}. For nanoshells with aspect ratios $\rho$ around 0.6, which we will be mostly concerned with, the symmetric resonance is the one necessitating the lowest gain level $\Gth$ to cross the lasing threshold. Again, a more general approach of the lasing properties of both resonances, with proper comparisons of their lasing thresholds and intensities etc., is left for future work.

Let us now illustrate our findings about the lasing threshold on a realistic example: we consider a nanoshell of external radius $a=10$~nm, internal radius 6 nm, and aspect ratio $\rho=0.6$. The shell is assumed to be made of silver, with the following parameters: $\hbar\omega_\text{p}=9.6$~eV, $\hbar\gamma = 0.0114$~eV and $\epsilon_\infty/\epsilon_0=5.3$. The core is made of silica ($\epsilon_\text{b}/\epsilon_0=2.1316$), doped with gain elements for which we set $\mu=10$~D, $\hbar\Delta= 2\hbar/\tau_2 =0.15$~eV (close to experimental linewidths observed for dyes), corresponding to $\tau_2\simeq 0.009 $~ps, and $\tau_1=5\,\tau_2$. We assume that the pump rate is much faster than spontaneous emission in the emitters (strong pumping), so that we take $\widetilde N = 1$. Finally, the external medium is taken to be water ($\epsilon_\text{e}/\epsilon_0=1.7689$). 

\begin{figure}[h!]
\centering
\includegraphics[width=0.7\columnwidth]{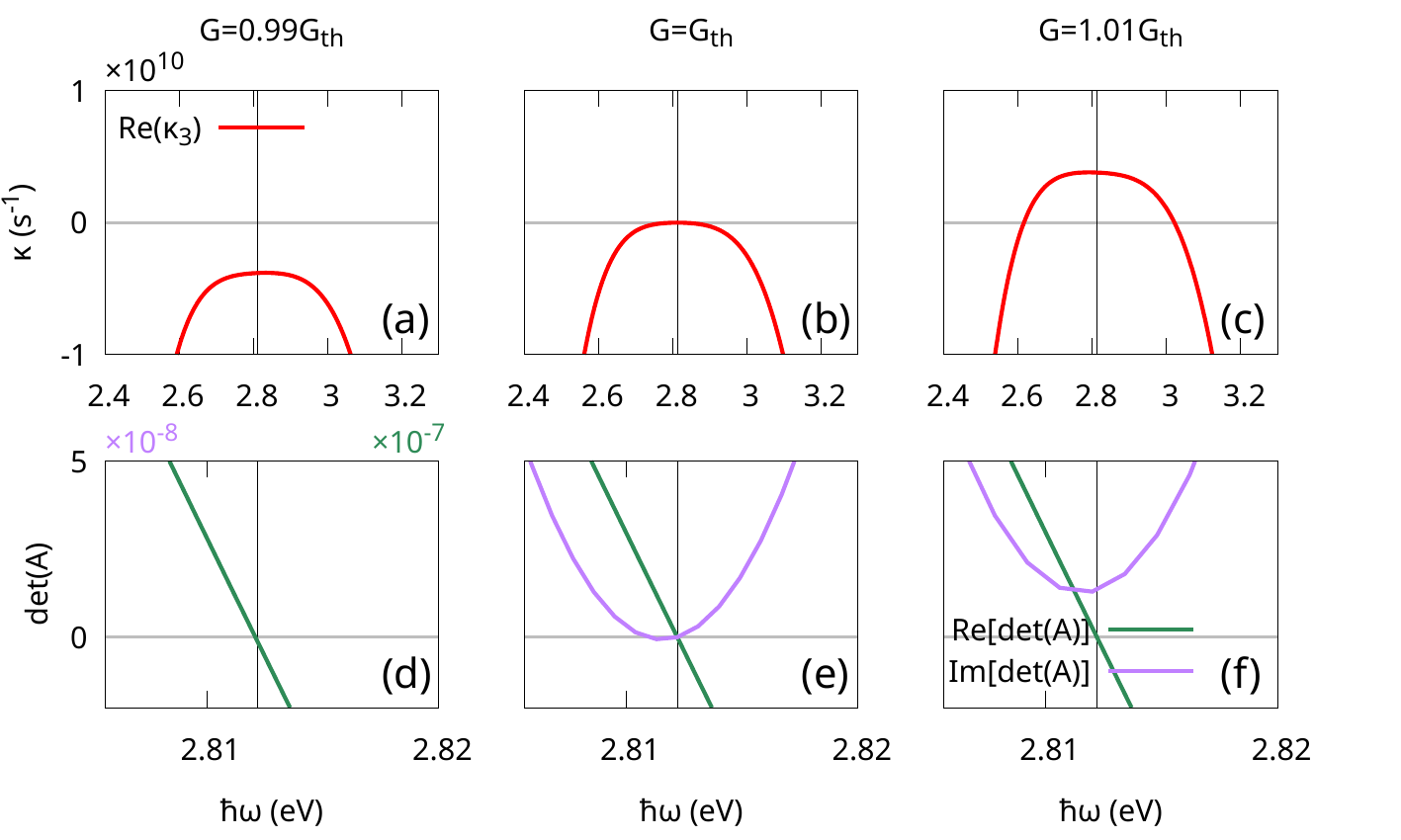}
\caption{(a-c): Plots of the real part of the eigenvalue $\kappa_3$ in the spectrum of the geometry matrix ${\bf A}$, respectively below, at, and above threshold. The real part of the eigenvalue is always strictly negative below threshold ($G<\Gth$), exactly null at threshold for $\hbar \omega=\hbar\omth\simeq 2.813$ eV and $G=\Gth\simeq 0.135$, and then strictly positive over a range of frequencies for $G>\Gth$. Other eigenvalues (not shown) always keep a negative real part.
(d-e): Real and imaginary parts of the determinant $\det({\bf A})$, respectively below, at and above threshold. The situation shown in (e), where both parts cancel simultaneously defines the value for the frequency of lasing at threshold $\omth=\omres$ and the gain value at threshold $\Gth$, according to eq~\ref{eq:lasingcond}. Vertical lines in all plots are guides for the eye showing the position of $\omth$. Parameters values are (see main text for explanations): $a=10$~nm, $\rho=0.6$, $\hbar\omega_\text{p}=9.6$~eV, $\hbar\gamma = 0.0114$~eV, $\epsilon_\infty/\epsilon_0=5.3$, $\epsilon_\text{b}/\epsilon_0=2.1316$, $\epsilon_\text{e}/\epsilon_0=1.7689$, $\mu=10$~D, $\hbar\Delta= 2\hbar/\tau_2 =0.15$~eV, $\tau_1=5\,\tau_2$, $\widetilde N = 1$.}
\label{fg:eigen}
\end{figure}

Solving eq~\ref{eq:denominatorcond} for the threshold conditions with the above numerical values, we find $\hbar \omth = \hbar \omres \simeq 2.813$ eV and $\Gth \simeq 0.135$. (And we take $\omg=\omth$ as per the optimal gain condition.) Figure~\ref{fg:eigen} shows the evolution of the eigenvalues of ${\bf A}$ when $G$ is increased for values below, at and above the threshold: it can be seen that the crossing of the lasing threshold is manifested by the appearance of a positive real part in one eigenvalue.

\begin{figure}[h!]
\centering
\includegraphics[width=0.7\columnwidth]{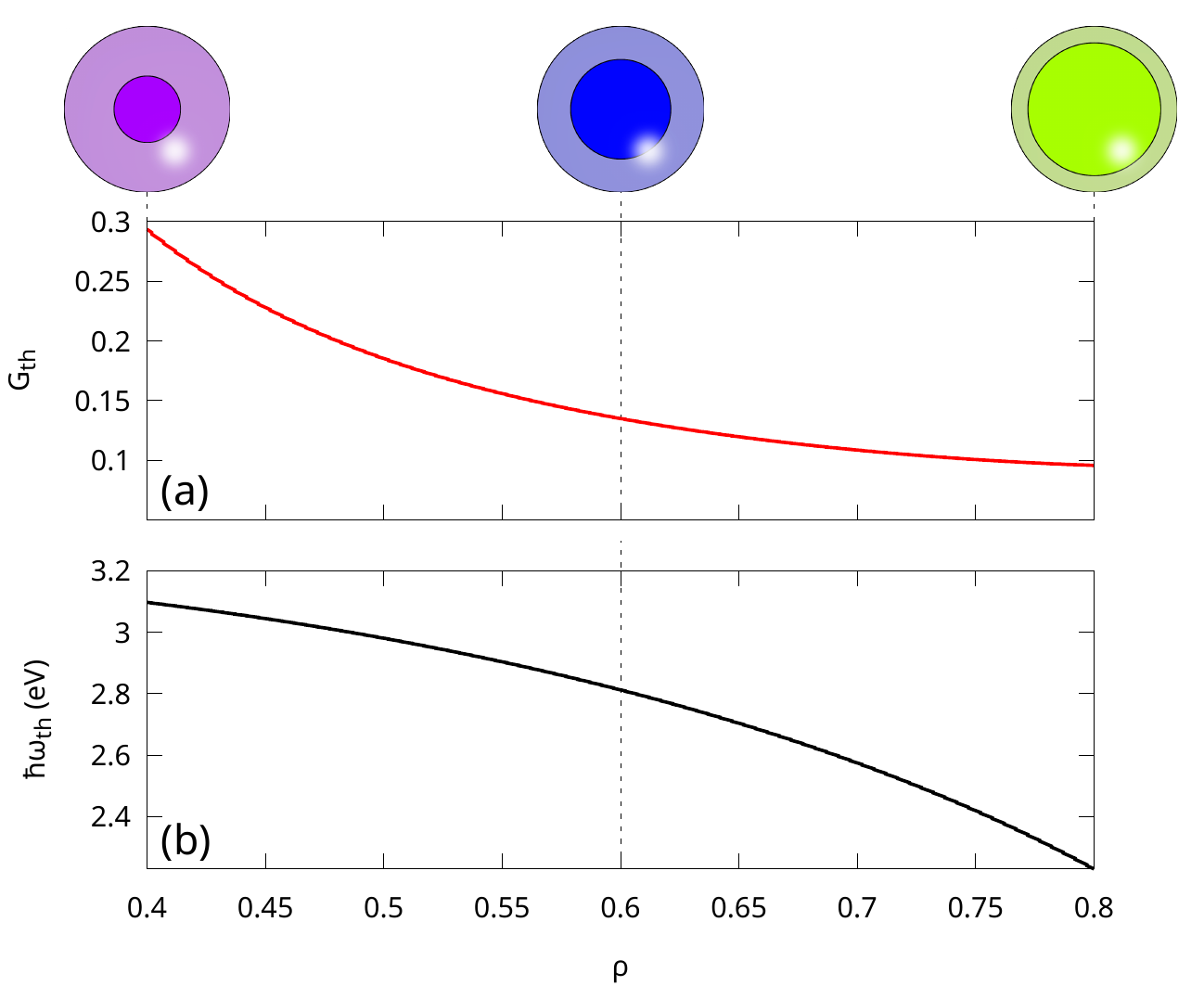}
\caption{Lasing threshold conditions for a gain-doped silver nanoshell, as a function of the aspect ratio $\rho$. 
(a) Threshold gain value $\Gth$; (b) Lasing frequency at threshold $\omth$. Colors on the nanoparticle drawings correspond to the lasing frequency at threshold for $\rho = 0.4, 0.6, 0.8$. Parameters other than $\rho$ are the same as in Fig.~\ref{fg:eigen}.
}
\label{fg:threshold}
\end{figure}

Figure~\ref{fg:threshold} shows changes in the gain and frequency at threshold when the aspect ratio $\rho$ of the nanoshell is varied from 0.4 to 0.8. Under optimal gain positioning, both $\omth$ and $\Gth$ are functions of $\rho$ only~\cite{Caicedo:2022}. As is known about the symmetric resonance of plasmonic nanoshells, when $\rho$ is increased (thinner shells), the resonance frequency $\omres$ redshifts, and so does the lasing frequency at threshold since they are equal according to eq~\ref{eq:omegares}. Simultaneously, the gain $\Gth$ required to cross the lasing threshold decreases, since the quantity of metal becomes less and less; lasing becomes easier as the Ohmic losses that need to be overcome are progressively reduced.

\section{Below the lasing threshold}
\label{sc:belowth}
We now consider the behaviour with time of the gain-doped nanoshell below the lasing threshold, i.e., $G<\Gth$. 

We place ourselves in the small-signal regime, where gain saturation is negligible and fields remain small with respect to $E_\text{sat}$, which guarantees that the r.h.s. of eq~\ref{eq:Nmastereq} can be approximated to zero. We assume that the nanoshell is excited with an incoming probe field $E_0$, taken as a harmonic plane wave with a definite frequency $\omega$ and constant magnitude.

\subsection{Time-dependent analysis}
For simplicity, let us start with considering a situation where the population inversion at the initial time $t_0$ is $N(t_0)=\widetilde N$: according to eq~\ref{eq:Nmastereq}, $N$ does not evolve through time, i.e., $N(t)=\widetilde N = \text{const}$. Then, ${\bf A} (N)={\bf A}$ is a constant matrix in time and the differential system~\ref{eq:qmastereq} is linear. The complete evolution of the system over time is given by the sum of exponentials of eq~\ref{eq:diffsol} (usually called ‘‘homogeneous'' solution), plus a constant term stemming from $\bf b$ (usually called ‘‘particular'' solution):
\begin{equation}
    {\bf q}(t)= \sum_{n=1}^{n=3} {\bf \hat q}_n e^{\kappa_n t} + {\bf q}_\text{part}.
    \label{eq:qsolution}
\end{equation}
The obvious particular solution is the constant vector:
\begin{equation}
    {\bf q}_\text{part}= - {\bf A}^{-1} {\bf b}.
    \label{eq:qpart}
\end{equation}
Since below threshold, all eigenvalues $\kappa_i$ of the matrix ${\bf A}$ have a strictly negative real part, the homogeneous part of the solution~\ref{eq:qsolution} represents an exponentially-decaying transient response. After it has vanished out, only the constant part ${\bf q}_\text{part}$ remains, which must represent the steady-state response of the nanoshell. This steady state is linearly related to ${\bf b}$ and therefore proportional to $E_0$. eq~\ref{eq:qpart} can be easily solved numerically, from which all steady-state values for the $p_i$ are found using eq~\ref{eq:p3nn}--\ref{eq:p0nn}, completing the solution by providing the value of all fields and polarizations in all domains of space. 

Now, let us assume that the initial value of the population inversion $N(t_0) \neq\widetilde N$, which is the general case: then $N(t)$ will have a dynamics of its own, evolving in accordance to~\ref{eq:Nmastereq} over a typical duration $\sim \tau_1$, until it reaches its final value $N=\widetilde N$. The transient evolution of the nanoshell is then modified slightly with respect to the previous case where we had $N(t)=\widetilde N$. But it can be proved that, because now $N(t)\leq \widetilde N$ at all times, the transient will still be quickly vanishing. Therefore, the final steady state remains unaffected and still given by eq~\ref{eq:qpart}.
\begin{figure}[h!]
\centering
\includegraphics[width=0.7\columnwidth]{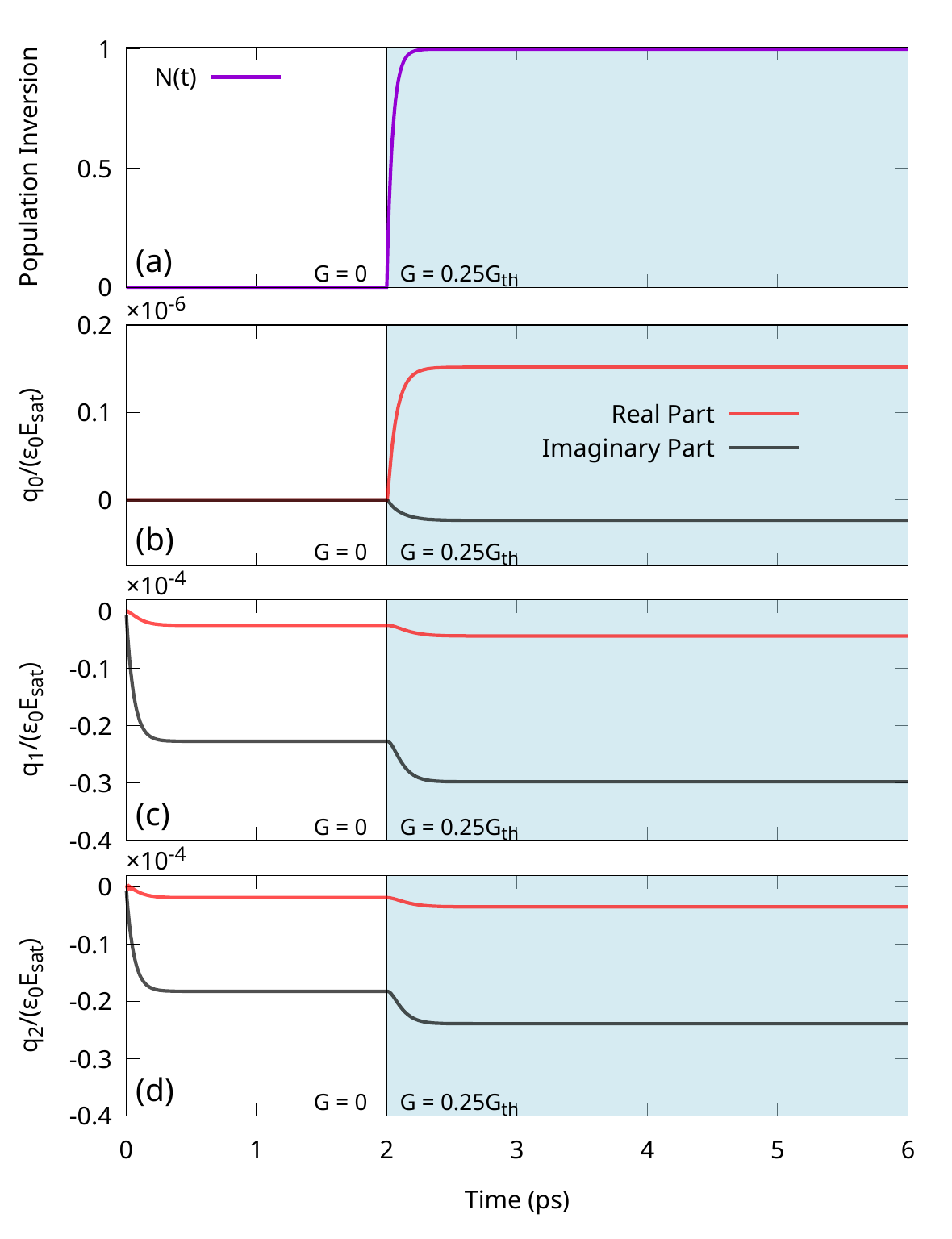}
\caption{Numerical solutions for the time evolution of the nanoshell's response below threshold, calculated at a fixed frequency $\hbar\omega=2.811$~eV, first with $G=0$ (no gain) for $t \leq 2$~ps, and then $G=0.25\,\Gth$ for $t > 10$~ps. (a)~Population inversion $N(t)$ versus time. Real and imaginary part of the dielectric polarization modes (b)~$q_0$; (c)~$q_1$; (d)~$q_2$ versus time. Parameters are the same as in Fig.~\ref{fg:eigen}.}
\label{fg:qblw}
\end{figure}

Figure~\ref{fg:qblw} illustrates this latter case of the time evolution of the nanoshell below threshold, with $N=N(t)$. We consider again the earlier example of a silver nanoshell with a gain-doped silica core (parameters are the same as in Fig.~\ref{fg:eigen}). In this case, $\Gth \simeq 0.1349$ and $\hbar \omth = \hbar \omres \simeq 2.8122$~eV, as obtained from the previous Section. We show the time evolution of the system computed from a numerical integration of eq~\ref{eq:qmastereq}--\ref{eq:Nmastereq}, for a fixed frequency $\hbar\omega=2.811$~eV. At $t=0$, the probe field $E_0$ is shone on the nanoparticle, which is assumed to initially have zero fields ($q_i=0$) and no population inversion ($N=0$). To start with, we assume $\widetilde N=0$ so that the gain value is set at $G=0$. The probe field magnitude is chosen very small to ensure that the systems stays well into the small-signal regime: $E_0 = 10^{-8} E_\text{sat}$. We see that after a short decaying transient due to eq~\ref{eq:diffsol}, all values for the field mode amplitudes  converge to their final values. This gives the stady-state response of the bare nanoparticle in the absence of gain, without any effect of the gain elements in the core. Then at $t = 2$~ps, conditions are changed: the gain value is set to $G=0.25 \Gth$ (sub-threshold level), and $\widetilde N$ is set to 1. It can be observed that $N(t)$ increases quickly from 0 to reach its final value $N=\widetilde N = 1$, while all other variables undergo a decaying transient evolution as explained above, converging to a new final value: this is now the steady state of the nanoparticle in the presence of gain. We note that all final values obtained in the presence of gain (for $t \gtrsim 3$~ps) are larger in absolute value than the ones without gain ($t \lesssim 2$~ps). This means that the response of the nanoparticle is amplified with the help of gain, as compared to the situation without gain.

Summarizing, our conclusions on the response of the gain-doped nanoshell below the lasing threshold are the following: \emph{(a)}~In the presence of an excitation of amplitude $E_0$, and in the small-regime signal, the response of the nanoshell is linear, proportional to $E_0$ and synchronized with it, i.e., oscillating with the same frequency $\omega$; \emph{(b)}~If there is no external excitation ($E_0=0$), the steady state of the nanoshell is null, i.e., there is no self-oscillation (as expected). These facts are often taken as granted in the literature on active nanoparticles below threshold. Our point here, however, is to lay out the proper mathematical justification supporting them, as they will soon be challenged when we switch to studying the situation above the lasing threshold.

\subsection{Steady-state polarizability}
The steady-state response of the nanoshell under external excitation can in fact be expressed in a much more familiar way, if one expresses the particle's external scattered field $p_3$ by solving eq~\ref{eq:p3nn}--\ref{eq:p0nn}. After some calculations (see SupplementaryInformation~\cite{SM} for details), one finds that $p_3$ is proportional to $E_0$:
\begin{equation}
    p_3= \frac{1}{4\pi a^3}\alpha E_0.
    \label{eq:p3alpha}
\end{equation}
where the polarizability $\alpha$ has the same classical expression as shown in eq~\ref{eq:polarizability}. In other words, this proves that under the threshold, the steady-state response of the nanoshell is simply given by the usual formula for polarizability, with the permittivites $\epsilon_\text{m}$ and $\epsilon_\text{g}$ given by their steady-state values from eq~\ref{eq:dru} and \ref{eq:lor}. By legitimating the use of the classical quasi-static polarizability, this also legitimates in retrospect the intuitive argument that the lasing threshold can be understood as the vanishing of the denominator of $\alpha$ in eq~\ref{eq:polarizability}, as was done in the previous section.

\begin{figure}[h!]
\centering
\includegraphics[width=0.7\columnwidth]{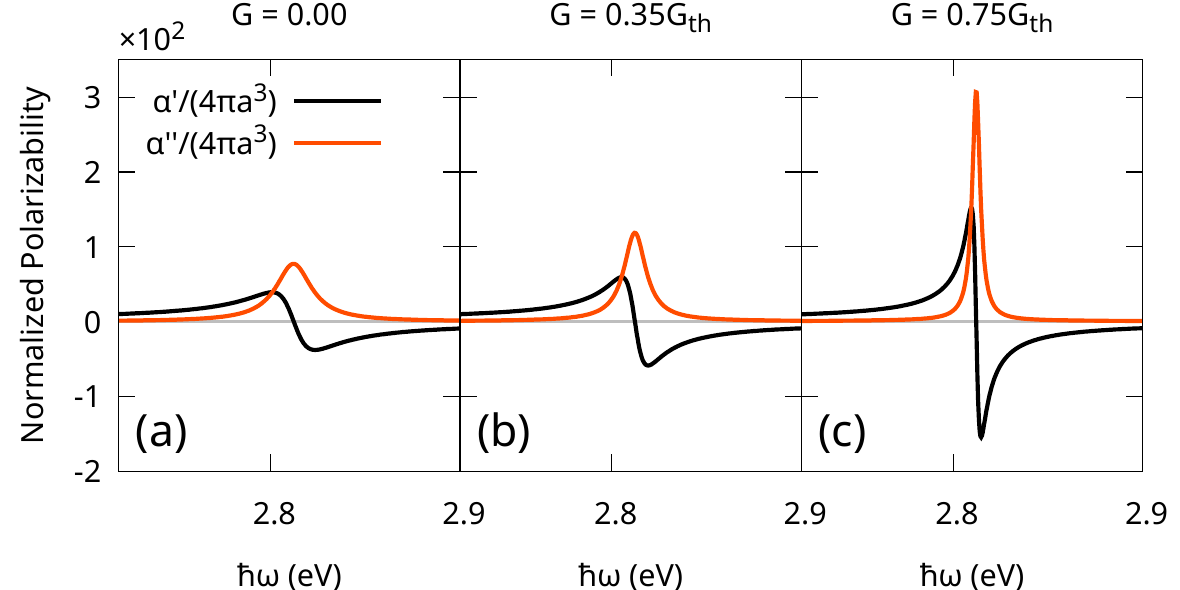}
\caption{Enhancement of the nanoshell polarizability $\alpha= \alpha' + i \alpha''$ for increasing values of the gain in the core. Parameters are the same as in Fig.~\ref{fg:eigen}.
(a) $G=0$ (no gain);
(b) $G=0.35 \,\Gth$;
(c) $G=0.75 \, \Gth$.
}
\label{fg:polblw}
\end{figure}

It is interesting to plot the evolution of the polarizability $\alpha(\omega)$ of the nanoshell for increasing values of the gain levels $G$ under the threshold, see Figure~\ref{fg:polblw}. It can be seen that below threshold, the effect of gain when it is increased, is simply to enhance the natural plasmon response of the nanoshell and improve its resonance quality.

This intuitive approach of describing gain-doped nanoparticles below threshold based on their quasi-static polarizability, is actually valid across other geometries beyond nanoshells (as long as a steady state with well-defined permittivities $\epsilon_\text{m}$ and $\epsilon_\text{g}$ exists in the final state, which is the generically true). Some of us had already demonstrated this fact in the case of a homogeneous, plasmonic sphere immersed in a gain medium~\cite{Veltri:2016}; calculations by us (not shown) also prove that this is true for core-shell nanoparticles with a metal core and gain shell. This justifies in hindsight all calculations that were made in ref~\citenum{Caicedo:2022} about gain-enhanced nanoparticles, based on this assumption. Note also that situations where nanoparticles are too large to be describable only with the quasi-static polarizability (i.e., multipolar modes are required) are studied in detail in ref~\citenum{Recalde:2023}.

Finally, we note that Equation~\ref{eq:p3alpha} is valid both in the small-signal regime that we studied, but also in the large-signal regime, when $\lvert E_\text{g} \rvert$ is comparable to $E_\text{sat}$. In the latter case, one needs to call upon the saturated permittivity~\ref{eq:slor} for the gain medium, making the polarizability $\alpha=\alpha(E_\text{g}$) nonlinear; Reference~\cite{Cerdan:2023} has a complete analysis of the behavior of a metallic nanoparticle with gain in this situation.

Therefore, below the lasing threshold, we conclude that, after a short-lived transient, the nanoshell simply acts as an ‘‘optical amplifier'', whereby the natural plasmon of the nanoshell can be significantly amplified and improved thanks to the assistance of optical gain. As this amplification regime was already studied in depth in a previous work~\cite{Caicedo:2022}, based on the simpler quasi-static polarizability approach, for various metals, geometries and aspect ratios, we need not delve into it further, and readily move to studying the lasing regime of the nanoshell when gain is raised above the threshold.

\section{Above the lasing threshold}
\label{sc:abvth}
We now turn to describing the situation of the nanoshell when it is pumped above its lasing threshold, i.e., when  $G > \Gth$.

\subsection{Time dynamics of the laser emission}

As explained in the ``Lasing threshold'' section, at threshold, an autonomous self-oscillation instability sets in, whereby fields become can become non-zero and start growing even in the absence of any exciting external probe field. The existence of such a sustained, autonomous self-oscillation state is indeed a central concept for lasers in general, and it is therefore crucial that its properties are fully studied and understood in the present case of lasing nanoshells. This is why, contrary to the previous section, we will here be concerned only with the \emph{free lasing state} of the nanoshell, i.e., when the probe field is zero ($E_0=0$), discarding any situations of external forcing. (Such situations will be briefly discussed in the conclusion of this article.)

Looking back at Fig.~\ref{fg:eigen}, we observe that when the gain level $G$ is that at threshold exactly, only the frequency $\omth=\omres$ is unstable, but when $G$ is increased above threshold, a wider and wider range of frequencies becomes unstable (i.e., there is a wider range where one eigenvalue of the matrix $\bf A$ has a positive real part). For example, when $G=1.01\,\Gth$, the unstable range lies approximately between 2.6 and 3 eV. Therefore, we expect the lasing instability to grow over a finite range of frequencies.

\begin{figure}[h!]
\centering
\includegraphics[width=0.7\columnwidth]{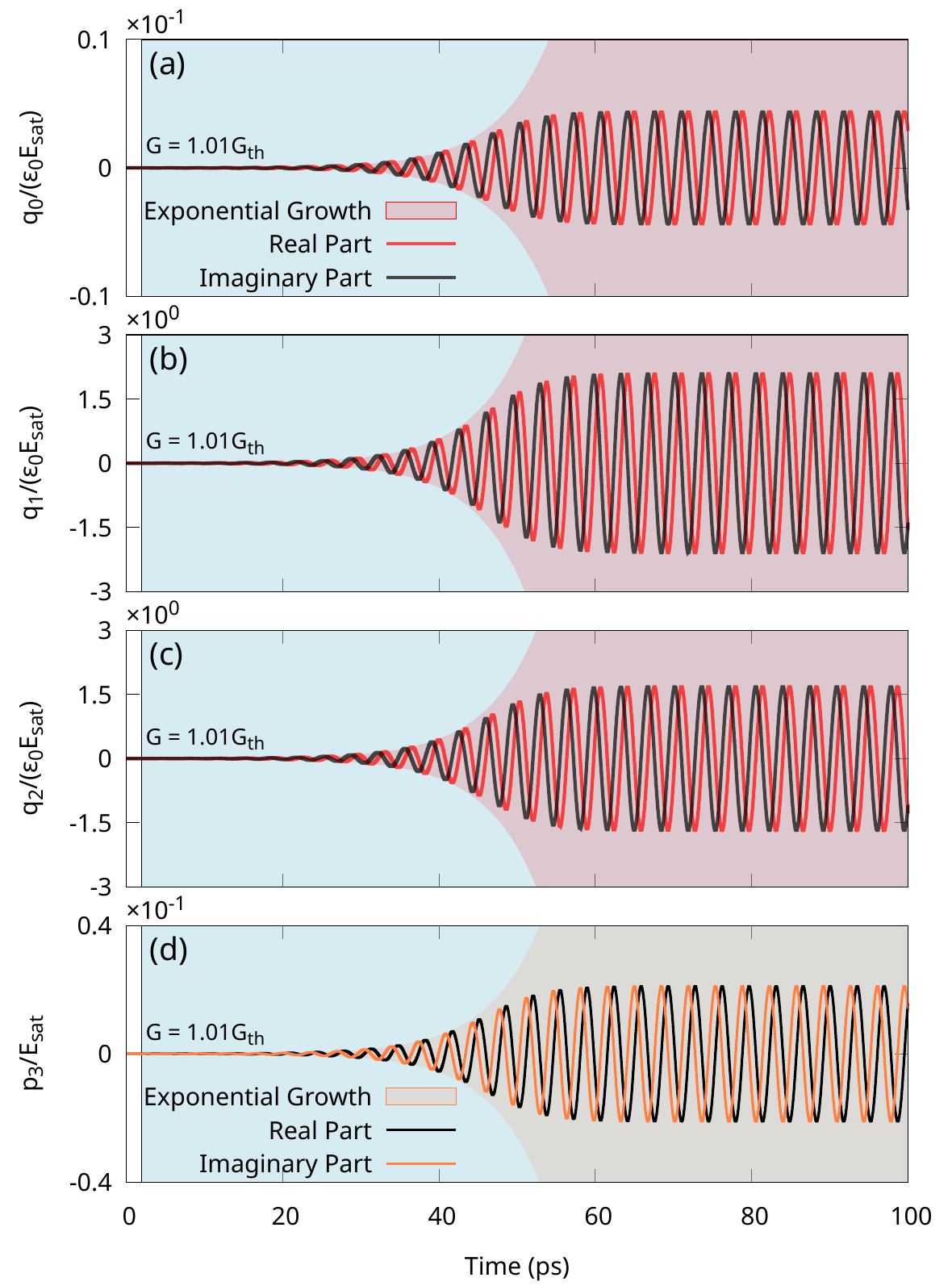}
\caption{Time dynamics of field and polarization amplitudes above the lasing threshold, for a gain value $G=1.01 \, \Gth$ and a frequency $\hbar\omega=2.811\text{~eV} < \hbar \omth$. A $t=0$, the gain is $G=0$, then it is set to $G=1.01 \,\Gth$ at $t=10$~ps. Displayed are the real and imaginary part of: (a) $q_0(t)$; (b) $q_1(t)$; (c) $q_2(t)$; (d) $p_3(t)$, normalized to $\epsilon_0 E_\text{sat}$ for polarizations and $E_\text{sat}$ for fields. Diverging envelopes correspond to the initial exponential growth computed from eq~\ref{eq:diffsol} when $N=\widetilde N$. Once gain saturation effects take place, the signal growth saturates and separates from the exponential envelopes, finally reaching a stable oscillatory state. Parameters are the same as in Fig.~\ref{fg:eigen}.
}
\label{fg:qabv}
\end{figure}

\begin{figure}[h!]
\centering
\includegraphics[width=0.7\columnwidth]{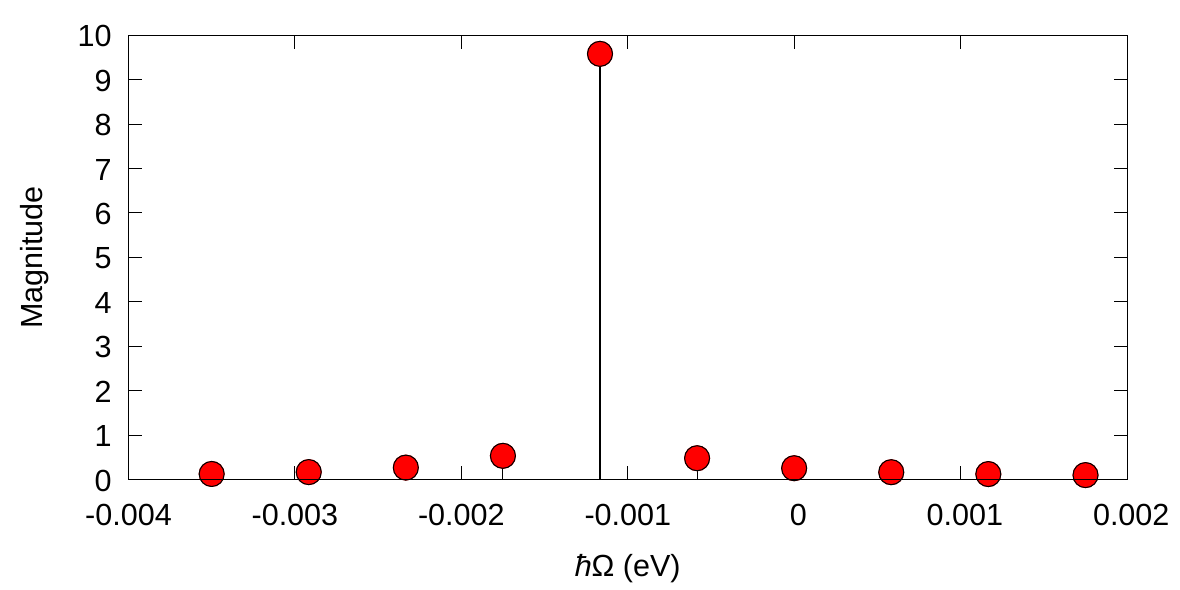}
\caption{Fourier analysis of $q_0(t)$ in the final oscillatory state shown in Fig.~\ref{fg:qabv}, yielding a single oscillation frequency $\hbar\Omega\approx -0.0012$~eV.  Similar analyses conducted on $q_1(t)$, $q_2(t)$, or $p_3(t)$, results in the same value for $\Omega$.}
\label{fg:fourier}
\end{figure}

To check this, and compare to the situation below threshold, we choose the same frequency as in Fig.~\ref{fg:qblw} ($\hbar\omega=2.811\text{~eV} < \hbar \omth$), and numerically solve the governing equations~\ref{eq:qmastereq}--\ref{eq:Nmastereq} for $G=1.01\, \Gth$. Results are shown in Figure~\ref{fg:qabv}.
(Computational details on how these results were obtained numerically can be found in the Supplementary Information~\cite{SM}.) We see that the variables $q_1$, $q_2$, $q_3$ first follow an exponential growth (led by the positive eigenvalue in $\bf A$); then this growth saturates and a stable long-term state is finally reached. We observe that this final state is \emph{oscillatory}, with all magnitudes $\lvert q_i (t)\rvert$ reaching a constant value; oscillations are purely sinusoidal with the same constant frequency $\Omega$ for all variables $q_i(t)$. The value of the frequency is obtained from a Fourier transform of the signals, see Fig.~\ref{fg:fourier}: $\hbar \Omega \simeq -1.2\times 10^{-3}$ eV. As before, from the $q_i(t)$, the field components $p_i(t)$ in all regions of space can be computed through the boundary conditions~\ref{eq:p3nn}--\ref{eq:p0nn}. Because these conditions are algebraic and linear, we find that all fields $p_i(t)$ follow the same temporal evolution as the $q_i(t)$, namely, an initial exponential growth and a saturation into a final oscillatory state with the same frequency $\Omega$: Figure~\ref{fg:qabv}-(d) displays the evolution of $p_3(t)$, which corresponds to the external field emitted by the nanoshell.

Thus, the nanoshell is able to maintain a stable emitted field $\lvert p_3\rvert \neq 0$, proving that it is indeed acting a nanolaser above the lasing threshold; and it is capable of doing so in an autonomous way, i.e., it reaches a free lasing state without any external excitation.
\begin{figure}[h!]
\centering
\includegraphics[width=\columnwidth]{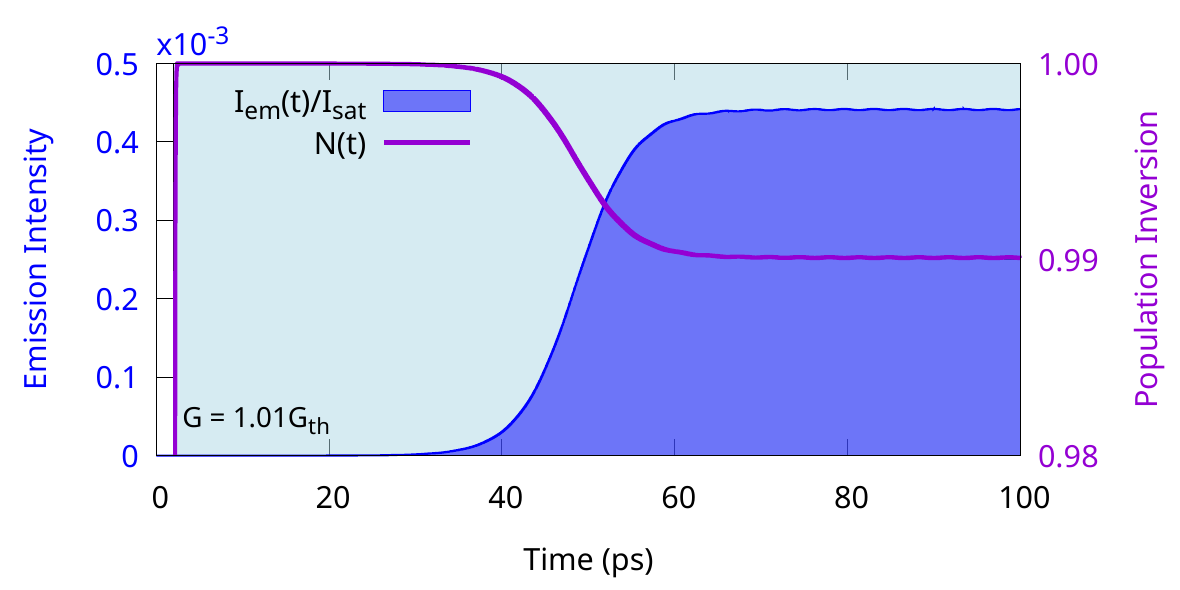}
\caption{Time dynamics of the population inversion and emitted intensity of the nanoshell above the lasing threshold, for a gain value $G=1.01 \, \Gth$ and a frequency $\hbar\omega=2.811\text{~eV} < \hbar \omth$ (same conditions as in Fig.~\ref{fg:qabv}). A $t=0$, the gain is $G=0$, then it is set to $G=1.01 \,\Gth$ at $t=2$~ps. Immediately, $N$ rises due to pumping, up to its maximal allowed value $N =\widetilde N = 1$. Then $N$ remains constant while the lasing instability slowly grows. Once the emitted intensity $I_\text{em}$ picks up, saturation terms come into play and decrease the value of $N$, until a steady state is finally reached for both quantities.}
\label{fg:abv}
\end{figure}
Figure~\ref{fg:abv} shows the intensity of emission radiated by the nanoshell in the lasing state $I_\text{em}(t)=\lvert p_3(t) \rvert^2$ at the same gain level and frequency ($G=1.01\,\Gth$ and $\hbar\omega=2.811$~eV), normalized to the saturation intensity $I_\text{sat}=\lvert E_\text{sat} \rvert^2$. The emitted intensity can be seen to pick up gradually and then reach a constant steady-state value, at the same time as the $q_i$ and $p_i$ reach their final oscillatory state. We also plot the evolution of the population inversion $N(t)$: initially, all fields are small and therefore, $N$ quickly comes close to $\widetilde N$. But as the emitted intensity $I_\text{em}$ increases, alongside with all fields inside the nanoshell, the saturation term $\text{Im}\!\left\{q_0p_0^*\right\}/(n\hbar)$ in eq~\ref{eq:Nmastereq} becomes significant, leading to a decrease in the value of $N(t)$ (corresponding to a depletion in the higher level of the two-level emitters due to stimulated emission). This decrease, in turn, limits the increase in the nanoshell's emitted intensity. Finally, an equilibrium is found and $N(t)$ stabilizes to a steady-state value where losses are exactly compensated by the pumping, as happens in a conventional laser. As expected (also from conventional lasers), this equilibrium is non-linear in nature due to the form of the saturation term $\text{Im}\!\left\{q_0p_0^*\right\}/(n\hbar)$.
\begin{figure}[h!]
\centering
\includegraphics[width=0.7\columnwidth]{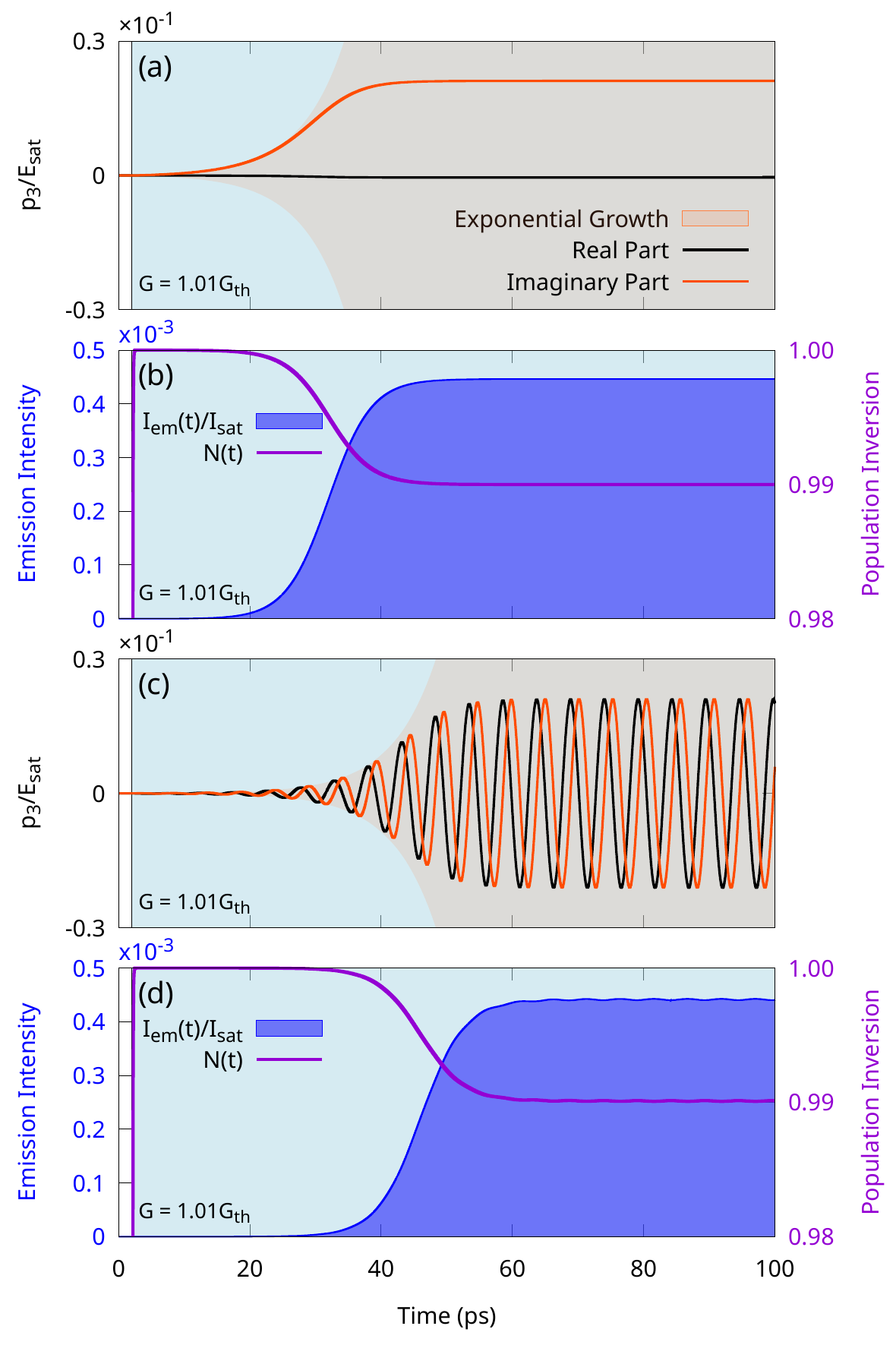}
\caption{Time dynamics of the emitted field $p_3(t)$, emitted intensity $I_\text{em}(t)=\lvert p_3(t) \rvert^2$ and population inversion $N(t)$ above the lasing threshold ($G=1.01 \, \Gth$) for two different frequencies: (a) and (b) $\hbar \omega=\hbar \omth=2.812$~eV; (c) and (d) $\hbar\omega=2.813\text{~eV} > \hbar \omth$. Parameters are the same as in Fig.~\ref{fg:eigen}.}
\label{fg:abvbis}
\end{figure}
Let us now explore other frequencies in the unstable range. Fig.~\ref{fg:abvbis} shows results for two other situations, for a frequency $\omega=\omth=\omres$ and for a frequency $\hbar\omega=2.813\text{~eV} > \hbar \omth$ (always keeping $G=1.01 \,\Gth$ as before). For both frequencies, all polarization and field modes $q_i(t)$ and $p_i(t)$ are found to follow the same generic trends as just before, i.e., an exponential growth saturating into a final state. (For conciseness, only the external field $p_3(t)$ has been plotted in the Figure.) For $\hbar \omega=\hbar \omth=\hbar \omres = 2.8122$~eV, no oscillations are observed, i.e., the final state has steady values with $\Omega(\omth)=0$, see Fig.~\ref{fg:abvbis}-(a) and (b). For $\hbar\omega=2.813\text{~eV} > \hbar \omth$, as shown in Fig.~\ref{fg:abvbis}-(c), the final state is found to be oscillatory again with a frequency $\hbar\Omega \simeq + 8\times 10^{-4}$~eV (obtained by Fourier transform, not shown).

We thus observe that the value of $\Omega$ depends on the frequency, i.e. $\Omega \equiv \Omega(\omega)$, with $\Omega \to 0$ when $\omega=\omth$, and a sign change when $\omega>\omth$ or $\omega<\omth$ (resp. $\Omega>0$ or $\Omega<0$).

In light of these findings, we can conclude that in the final lasing state above threshold, the variables in the system take on the following generic form:
\begin{align}
   q_i(t)&=q_i^\text{ss}\, e^{i\Omega t}, \label{eq:qfinal}\\
   p_i(t)&=p_i^\text{ss}\, e^{i\Omega t}, \\ 
   N &=N^\text{ss}, \\
   I_\text{em}&=I_\text{em}^\text{ss}=\lvert p_3^\text{ss}\rvert^2,\label{eq:Ifinal}
\end{align}
where all quantities with the superscript `ss' for ``steady state'' are constants and the frequency $\Omega$ depends on $\omega$. From~\ref{eq:qfinal}, we deduce that the vector ${\bf q}=[q_0, q_1, q_2]^\mathrm{T}$ defining the electromagnetic state of the system, as introduced in eq~\ref{eq:qdef}, has the following form in the final state:
\begin{equation}
    {\bf q}(t) = {\bf Q} e^{i \Omega t}, 
    \label{eq:qdependence}
\end{equation}
where ${\bf Q}$ is the constant vector $[q_0^\text{ss}, q_1^\text{ss}, q_2^\text{ss}]^\mathrm{T}$.

\subsection{Steady-state lasing with a shifted frequency}

To interpret properly the meaning of the final oscillatory state for ${\bf q}$ as expressed in eq~\ref{eq:qdependence}, one needs to remember that all time-dependent variables were defined as slowly-varying enveloppes upon a $e^{-i \omega t}$ carrier wave [see eq~\ref{eq:rea}]. Therefore, the complete time dependence of electrical fields and polarizations is proportional to ${\bf q}(t) e^{-i \omega t}$, which, using eq~\ref{eq:qdependence}, writes as

\begin{equation}
    \label{eq:freqshift}
    {\bf q}(t) e^{-i \omega t} = {\bf Q} e^{-i (\omega - \Omega) t}={\bf Q} e^{-i \omem t},
\end{equation}
where we have defined a new frequency $\omem$ corresponding to the frequency $\omega$ shifted by an amount $\Omega$:
\begin{equation}
    \label{eq:omprime}
    \omem\equiv \omega - \Omega.
\end{equation}

The last equality in eq~\ref{eq:freqshift} is of high physical significance: it demonstrates that in the final lasing state, all fields and polarizations, including the electrical field emitted by the nanoshell, are purely sinusoidal \emph{with a frequency} $\omem$ (we recall that ${\bf Q}$ is a constant). Hence the nanolaser really emits at the shifted frequency $\omem=\omega - \Omega$, not at the frequency $\omega$ of the carrier wave. Furthermore, this lasing state is a true steady state in the sense that all physical quantities (fields, polarizations intensities, and population inversion) are constant in time at frequency $\omem$, due to the constant value of ${\bf Q}$.

We conclude that the oscillations seen in the final state of the variables $q_i(t)$ and $p_i(t)$ were apparent ones, when observed relatively to the carrier wave of frequency $\omega$; but once all time dependences are taken into account, the physical final state of the nanoshell is indeed one of constant steady-state lasing with frequency $\omem$.

It is important to clarify that the initial frequency $\omega$ has no physical meaning intrinsically: it is just the frequency around which the rotating wave approximation has been taken to write the differential system of eq~\ref{eq:1ddn}--\ref{eq:3ccn}. Since there is no externally imposed probe field associated to this frequency $\omega$ and we consider situations where the nanoshell is left to freely oscillate above the lasing threshold (self-oscillation), the resulting frequency of emission has no particular reason to be the same as the arbitrary $\omega$. Here, $\omega$ shall simply be considered as a mathematical parameter in the differential system, which can be varied continuously so as to scan the full range of emission of the nanoparticle; the actual physical frequency of the nanolaser emission corresponding to each chosen $\omega$ can be calculated as $\omem=\omega - \Omega$, where $\Omega(\omega)$ is found as part of the solution of the differential equations.
\section{Maximal emission spectrum of the nanolaser}\label{sc:es}
With the help of the results obtained in the previous Section, we are now in a position to build the spectrum of emission of the nanoshell laser above its lasing threshold.

The procedure to calculate the spectrum of emission, is to compute the whole set of emission frequencies $\omem=\omega - \Omega$ of the nanoshell, by scanning the value of the parameter $\omega$ over the unstable range and finding the value of $\Omega(\omega)$ as explained above; then, for each these value found for $\omem$, one calculates the value of the steady-state intensity of emission $I_\text{em}(\omem)=|p_3^\text{ss}|^2$ according to eq~\ref{eq:Ifinal}, by picking the final value $p_3^\text{ss}$ as obtained from the corresponding numerical solution. 

We emphasize that the intensity $I_\text{em}(\omem)$ obtained for any $\omem$ frequency is here calculated by taking each of them separately from all others, i.e., without considering the effect of simultaneous emission in other $\omem$ frequencies. In other words, we are assuming that each frequency is able to lase independently from others, and is free to use all of the available gain brought by external pumping at will, without competition from other frequencies. To which extent this assumption may hold true will be considered in the ``Discussion'' section; what can be said for certain at this stage is that under this independency assumption, all $\omem$ frequencies in the spectrum are able to lase up to their full capacity--while all frequencies outside this spectrum cannot lase at all (since they are not unstable by self-oscillation). Therefore, the spectrum that we calculate here really represents the \emph{maximal spectrum of emission of the nanolaser}, in the sense of the widest possible that can be expected, with maximal possible intensities for all unstable frequencies.

\begin{figure}[h!]
\centering
\includegraphics[width=\columnwidth]{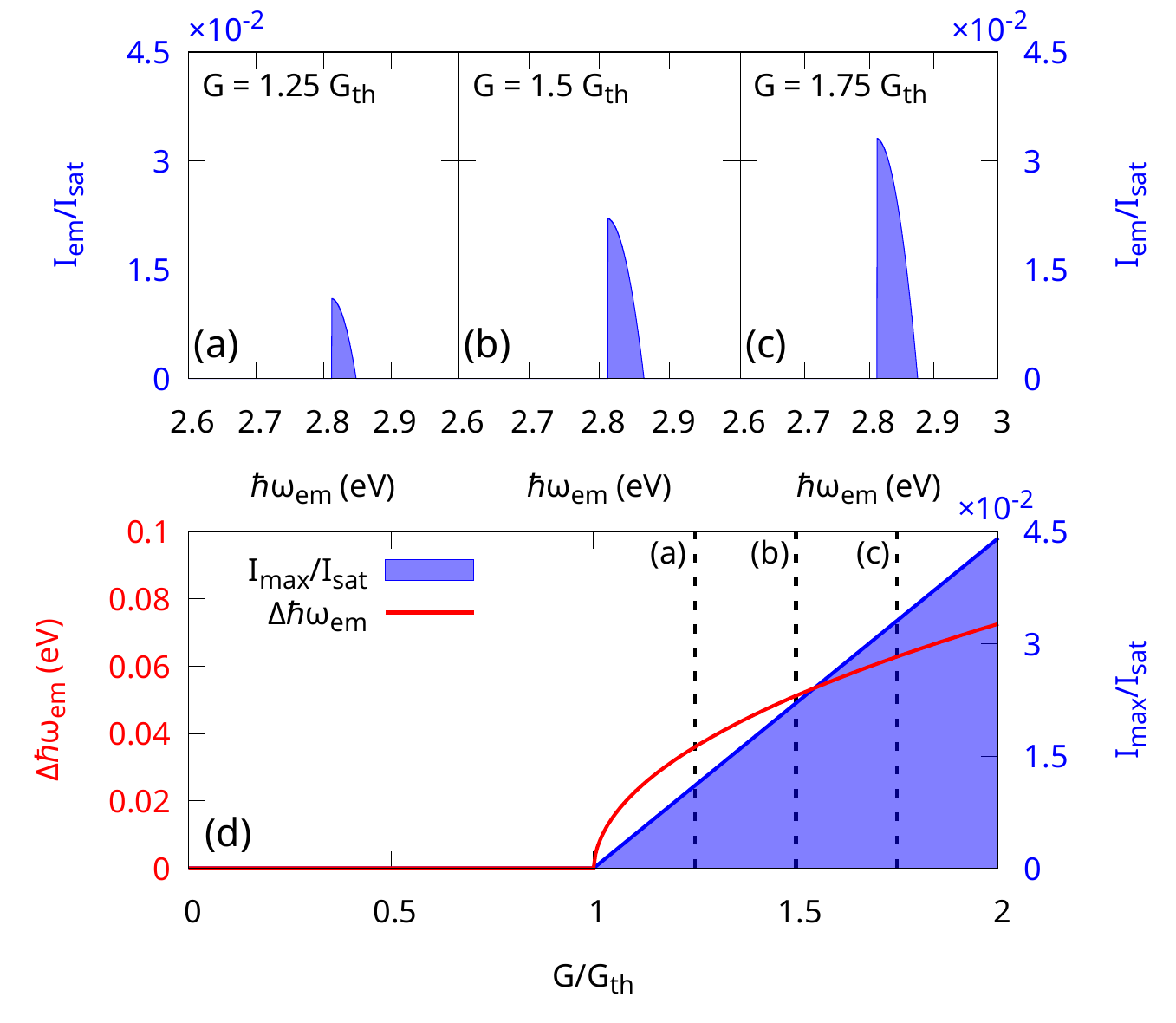}
\caption{Emission spectrum of the nanoshell $I_\text{em}$ as a function of the emission frequency $\hbar \omem$ in the lasing steady state, for increasing gain levels: (a) $G=1.25 \,\Gth$; (b) $G=1.5 \,\Gth$; (c) $G=1.75 \,\Gth$. (d) Emission linewidth $ \Delta \hbar\omem$ and peak intensity $I_\text{max}$ as a function of gain. Vertical dashed lines mark the points corresponding to the spectra shown in (a), (b), and (c). The linewidth is taken as the full width measured at the base of the emission spectrum. Parameters are the same as in Fig.~\ref{fg:eigen}.}
\label{fg:Grange}
\end{figure}

In Figure~\ref{fg:Grange}-(a)--(c), we plot this maximal spectrum of emission $I_\text{em}(\omem)$ as obtained from the procedure exposed above, for a nanoshell with aspect ratio $\rho=0.6$ and for three increasing gain levels: $G=1.25\,\Gth$, $G=1.50\,\Gth$, and $G=1.75\,\Gth$. 
One observes that the emission lines are relatively thin, and asymmetric in shape, with the intensity peak located at the lower-frequency edge [note that this holds under the condition of optimal gain positioning, see eq~\ref{eq:optimalgain}]. We also find that this sharp, lower-frequency edge of the emission line (with the associated intensity peak) corresponds precisely to the lasing frequency at threshold $\omth=\omres$.

We therefore come to the following striking conclusion: due to the frequency displacement $\Omega$, the resulting spectrum of emission is one-sided with respect to the plasmon resonance, i.e., the emission is strictly restricted to frequencies such that
\begin{equation}
\label{eq:gtrthanomres}
   \omem \geq \omres.  
\end{equation}
The numerical reason for this is because the values we compute for $\Omega(\omega)$ are always such that $\Omega\leq\omega-\omres$; which means that, in accordance with eq~\ref{eq:omprime}, we always have $\omem - \omres \geq 0$ for emission frequencies.

Finally, without much surprise, as the gain level is increased from Figure~\ref{fg:Grange}-(a) to (c), we see that the intensity of the nanolaser emission increases significantly, and that the range of emission widens as well. However, all in all, the aspect ratio of the emission band remains globally similar, meaning that the quality of the nanolaser emission does not depend much on the level of gain provided to the system.

Figure~\ref{fg:Grange}-(d) shows the evolution of the maximum (peak) lasing intensity in the spectrum, $I_\text{max}$, and the linewidth of the emission, $\Delta \hbar \omem$, as functions of the gain level $G$ normalized to the gain threshold $\Gth$. As expected, the emission appears when $G/\Gth=1$. From that point onward, both the maximum emission intensity and the width monotonously increase. We observe in particular that the increase of $I_\text{max}$ vs. $G/\Gth$ is strictly linear. It is interesting to consider the typical values obtained for the nanolaser linewidth: the observed values for $\Delta\hbar\omem$ are of the order of a few $10^{-2}$~eV, which correspond to a linewidth of a few nanometers; for example, for $G=1.5\,\Gth$, we have $\Delta\hbar\omem\simeq 0.05$~eV, which gives a linewidth of around 8~nm. (Let us recall, as explained earlier, that this linewidth should be considered as the \emph{widest} possible to be expected from the nanolaser, see the ``Discussion'' section.)

\begin{figure}[h!]
\centering
\includegraphics[width=0.7\columnwidth]{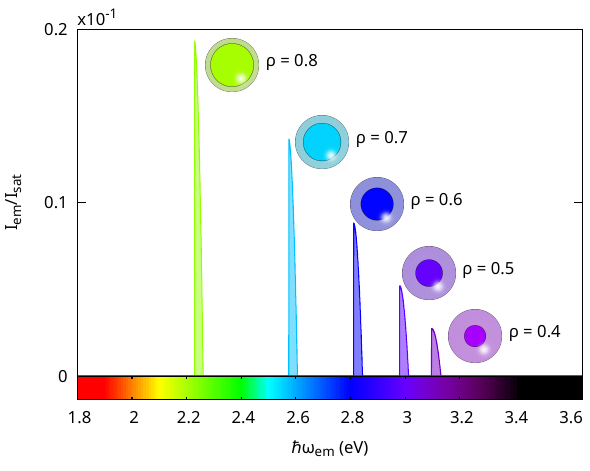}
\caption{Emission spectra $I_\text{em}(\omem)$ of lasing nanoshells for varying aspect ratios $\rho$, with a gain level set at $G=1.2 \,\Gth$, and normalized to $I_\text{sat}$. Nanoshell drawings bear the actual colors corresponding to the emission spectrum. All parameters beside $\rho$ are the same as in Fig.~\ref{fg:eigen}.}
\label{fg:rho}
\end{figure}

One well-known feature of nanoshells is that the position of their plasmonic resonances can be easily tuned by changing the thickness of the metallic shell (i.e., by changing the nanoparticle's aspect ratio $\rho$): how does this reflect in the emission spectrum of the nanolaser? Figure~\ref{fg:rho} shows the evolution of the maximal emission spectrum as a function of $\rho$, when the shell size is modified, keeping the external radius of the nanoparticle constant ($a=10$~nm). One can see that the spectra are indeed strongly dependent on $\rho$ and cover most of the visible region, from green ($\rho=0.8$) to violet ($\rho=0.4$). Hence, the specific nanoshell geometry indeed makes a versatile choice for applications.

We also note that the peak intensity of the emission increases strongly as $\rho$ is increased (from violet to green). This is because larger $\rho$ values correspond to thinner metallic shells, and therefore to smaller associated Ohmic losses responsible for dampening the emission. If we compare absolute gain levels, taking into account the values found earlier for $\Gth$ (see Fig.~\ref{fg:threshold}), for the violet emission ($\rho=0.4$), we have an absolute gain value $G=1.2 \,\Gth\simeq 0.35$, while for the green emission ($\rho=0.8$), we have $G=1.2 \,\Gth \simeq 0.11$. In other words, a green-lasing nanoshell is much more efficient than a violet one, as it delivers an emission seven times more intense with a quantity of gain more than three times smaller.

\section{Summary}
We studied the emission and lasing properties of a nanoshell particle made of an externally pumped, active core and a plasmonic shell, with the help of a set of space and time-dependent governing equations. These coupled equations are, on one hand, the Drude equation of motion for the free electrons within the metallic part, and the other hand, the optical Bloch equations, accounting for population changes occurring between the two electronic levels in the gain material part. In the nanoshell geometry specifically, the dipolar mode is the only one to be excited and emitting, including in the lasing regime. Therefore, we used a quasi-static description for fields based on an expansion in spherical harmonics, keeping only dipolar terms. The set of governing equations was then projected onto these dipolar terms and put under matrix form to allow for numerical solving.

With the help of a linear instability analysis of the governing equations, we first demonstrated the existence, then calculated the value, of the gain threshold value $\Gth$ above which the nanoshell hosts a self-oscillation instability (i.e., the emission of a lasing field in the absence of any exciting probe field). Below this threshold, no instability exists and the nanoparticle can only react to an external excitation. 

We then studied the situation prevailing when the gain level is lower than the threshold ($G<\Gth$). When submitted to a probe field, the nanoparticle produces transient fields which rapidly decay to a steady-state response, which is linearly proportional to the amplitude of the exciting probe field $E_0$. One may then use the standard quasi-static formula for the polarizability of a nanoshell, eq~\ref{eq:polarizability}, to calculate the dipolar moment of the particle relative to $E_0$, with the help of the gain-dependent permittivity of eq~\ref{eq:lor} or \ref{eq:slor}. In this regime, the nanoshell acts as a plasmonic amplifier, i.e., it synchronizes and responds linearly to the external field, and its response becomes more and more intense and sharp as more gain $G$ is provided.

Next, we studied in depth the lasing regime above the threshold ($G>\Gth$). We exclusively considered autonomous situations, where the nanoshell oscillates freely in the absence of any externally-imposed field. The self-oscillation of the nanoshell initially grows exponentially, according to the results of the linear instability analysis made previously. After a while, stimulated emission due to the lasing process starts exceeding the capacity of the pump, effectively reducing the population inversion (saturation effect) and limiting the intensity growth. This brings the particle to a final state of steady-state emission. 

This final lasing state is characterized by the following salient features:

\emph{(i)} The determination of the actual values of the emission wavelengths $\omem$ within the spectrum requires extra care, since the nanolaser is free to choose its lasing frequency. The actual wavelength of emission $\omem$ is given as $\omem = \omega - \Omega(\omega)$, where $\Omega(\omega)$ is a frequency shift (frequency-pulling effect) from the rotating-wave frequency $\omega$. The shift $\Omega$ is computed by Fourier analysis from the numerical steady-state solution.

\emph{(ii)} For a given nanoshell, the range (spectrum) of emission wavelengths $\omem$ depends on the applied gain $G$. When $G$ is increased, and the threshold is crossed, the lasing start at a single frequency $\omth=\omres$, where $\omres$ is one of the nanoparticle's plasmon resonance frequencies. Then, as $G$ is further increased above threshold, the emission range $\Delta\hbar\omem$ widens. We calculated the corresponding maximal (widest possible) spectrum of emission of the nanolaser and find typical linewidths in the range 5--10 nm. Simultaneously, we find that the peak lasing intensity $I_\text{max}$ increases linearly as the level of gain $G$ is increased.

\emph{(iii)} Due to the action of the frequency shift $\Omega$ moving unstable frequencies around, we observe that the final (maximal) spectrum of emission of the nanolaser is one-sided with respect to the plasmon resonance frequency $\omres$, that is, the lasing occurs only for frequencies $\omem \geq \omres$.

\emph{(iv)} The color of the nanolaser emission can be tuned in a versatile way across the visible range by choosing nanoshells of various aspect ratios $\rho$. Nanoshells with thinner metallic shells (emitting on the low-energy end of the visible) are more efficient, i.e., much more intense with less required gain, than those with thicker shells (emitting on the high-energy end of the visible).

\section{Discussion}
The results summarized just above now require some thorough discussion to assess their physical significance and validity.

Regarding our results on the determination of the lasing threshold and the amplification regime below this threshold: they globally confirm earlier similar findings in the literature, justifying in particular the common use made of the standard polarizability formula and of steady-state permittivities for the gain material (whether linear or saturated).

Let us now turn to the lasing regime, above threshold. It is the first time, to our knowledge, that the nonlinear lasing steady state of a nanoshell, alongside with the dynamics leading to it, has been fully characterized. It is also the first time that the maximal spectrum of emission of a lasing nanoshell has been calculated.

Several earlier works have studied the nanoshell geometry in the lasing regime, mostly focusing on driven situations where the nanoshell is under the action of a probe field. Authors of Refs.~\cite{Baranov:2013,Arnold:2015} in particular have briefly considered the autonomous (free) situation, writing equations closely similar to our set of eq~\ref{eq:1ddn}--\ref{eq:3ccn}. They looked for the steady-state regime of lasing using the steady-state expression of the permittivity $\epsilon_\text{g}(\omega)$ in the gain region, including saturation effects, eq~\ref{eq:slor}, and concluded that autonomous lasing can only occur at one single frequency $\omega=\omres$, equal to one of the plasmonic resonance frequencies of the nanoshell, for all gain levels above the threshold. Our findings show that this line of reasoning is incomplete. We also find that lasing at one single frequency occurs only when the gain level is set right at the threshold value, but then the spectrum of emission widens as $G$ is increased above the threshold, with a finite linewidth $\Delta\hbar\omem$. The reason why this fact was overlooked is because the use of a steady-state permittivity $\epsilon_\text{g}(\omega)$ in the lasing regime is incorrect; steady-state permittivies follow from cancelling all time derivatives in eq~\ref{eq:1ddn}--\ref{eq:3ccn}, or equivalently, in eq~\ref{eq:qmastereq}--\ref{eq:Nmastereq}. However, we have found that the final state of lasing is a steady state with a shifted frequency with respect to the frequency $\omega$ used in the rotating-wave approximation. This means that, when expressed in terms of a carrier wave at frequency $\omega$, field amplitudes show an extra oscillation at frequency $\Omega$, and therefore the aforementioned time derivatives are \emph{non-zero} (except for the equation on $N$). The only frequency where there is no extra oscillation (i.e., where $\Omega=0$ and time derivatives do cancel) is $\omem=\omres=\omth$, as seen in Fig.~\ref{fg:abvbis}-(a). Therefore, the conclusions about the free lasing regime of nanoshells as written in refs~\citenum{Baranov:2013,Arnold:2015} do only apply to that specific frequency but miss out on the rest of the spectrum. We emphazise, however, that the authors of ref~\citenum{Baranov:2013} have correctly predicted the linear dependence of the peak intensity $I_\text{max}$ versus the gain level $G/\Gth$ seen in Fig.~\ref{fg:Grange}-(d), because that peak intensity indeed occurs at the specific frequency $\omem=\omres$ (under conditions of optimal gain positioning). Beyond these considerations on the free lasing regime, to what extent the existence of other emission frequencies $\omem\neq \omres$ should also modify conclusions drawn in refs~\citenum{Baranov:2013,Arnold:2015} about the \emph{driven} (forced) regime of oscillation, is an open question at this stage.

One striking and novel result of our study is that the emission of the nanolaser is strongly asymmetrical, since it only occurs on the high-frequency side of the nanoshell's natural plasmon resonance ($\omem \geq \omres$). This is due to the effect of the frequency shift $\Omega$ (also known as a frequency pull-out) which translates and ‘‘folds'' the initially symmetric unstable range of Fig.~\ref{fg:eigen} to that one side only of the plasmon resonance. (Note that we did not find numerically any situation where this pull-out effect would result in a spectrum located on the opposite, lower frequency-side of $\omres$.) We are not aware, to the best of our knowledge, of any similar claim made explicitly in the existing literature on nanolasers, either theoretically or experimentally. Let us underline that it is uncertain whether this phenomenon would extend to other geometries than nanoshells, and that in any case, exhibiting this effect experimentally would be challenging, since most experiments are made on collections of individual nanolasers: any statistical dispersion in the structural properties of the nanoresonators will certainly smear out the asymmetry of the emission line. We note, however, the existence of sharply asymmetric emission spectra on the high- or low-frequency side for example in refs~\citenum{Vass:2024,Lu:2014}.

Finally, a discussion is due on the actual spectrum of emission to be expected from the nanoshell laser. In our study, we have calculated this spectrum under the assumption of independent emission of the various frequencies $\omem$ composing it, meaning that all frequencies are allowed to consume energy from the two-level system as if they were alone. They would therefore all grow to their maximal capacity, which is why we called our calculated spectrum ‘‘maximal'', i.e. the widest one with all frequencies emitting to the highest possible intensity. However, this independency hypothesis is clearly incorrect because all $\omem$-frequencies in fact draw energy from \emph{the same reservoir} of excited electrons represented by the population inversion $N$. Whatever is consumed by one frequency (making $N$ decrease), is not available to another one, and thus these  ‘‘modes'' are truly competing for the same energy resource. This is because the dispersion in the $\omem$-frequencies originates in the finite width of the gain curve feeding the nanolaser, as illustrated by the Lorentzian curve of eq~\ref{eq:lor}, which is well-known in the literature on classical (macroscopic) lasers as a situation is known of \emph{homogeneous broadening}. In such a situation,  it is established  that the numerous laser modes inside the initially unstable range of the spectrum will indeed compete, and only the mode with the fastest growing rate will ultimately survive---usually the first mode to reach the threshold (see for example Chap.~8 in ref~\citenum{Verdeyen:1995} or Chap.~11 in ref~\citenum{Siegman:1986}). This winning mode will then sharpen by several orders of magnitude, as it remains alone in the cavity and may consume all the available gain inside it at will, until reaching some final limit of acuteness which will be discussed below. Therefore, the actual width of emission of a homogeneously broadened laser is set by the ultimate width of this surviving mode only, not by the initial width of the gain curve or anything of that order.

If we were to apply this line of thought in our case, this would mean that only the frequency at $\omem=\omres$ (where the intensity is maximal) will eventually survive in the final lasing state of the nanolaser, eliminating all other $\omem$ frequencies, resulting in a width of emission possibly much thinner than the spectra presented in Figs.~\ref{fg:Grange} and \ref{fg:rho}. In the present stage, it is unclear to what extent, if any, this scenario should apply here. Several qualitative facts should indeed be taken into consideration. Firstly, the classical scenario applies well to situations with spectrally well-defined cavity modes, much thinner than the unstable gain curve, whereas we are here in presence of several frequencies acting within a rather wide mode (the dipolar plasmon resonance) inside an unstable range. Secondly, the curve of the growth rates for unstable frequencies, given by $\text{Re}(\kappa_3)$ in Fig.~\ref{fg:eigen}, is very flat around its maximum: therefore, it may be that the contrast between the central frequency $\omth=\omres$ and neighbouring ones is not significant enough for the former to dominate the competition. Thirdly, in the classical homogeneously broadened laser scenario, the final surviving mode controls the final width of emission because it is able to become much thinner than the homogeneously broadened gain transition. The well-known Schawlow-Townes formula~\cite{Verdeyen:1995,Siegman:1986} actually expresses the theoretical value $\Delta\omem$ of the limiting linewidth of the surviving mode, in the most ideal, case as
\begin{equation}
\label{eq:ST}
   \Delta\omem \sim \hbar\omem \frac{\Delta{\omega_\text{cav}}^2}{P_\text{out}}.  
\end{equation}
where $\omem$ is the central emission frequency of the mode,  $\Delta\omega_\text{cav}$ is the width of the passive resonant cavity mode at the origin of the surviving mode, and $P_\text{out}$ the power output of the laser. In the case of a classical laser, we may typically have $\Delta\omega_\text{cav} \simeq 1~\text{MHz} \simeq 10^{-8}~\text{eV}$ and $P_\text{out} \simeq 1$~mW, which results in $\Delta\omem \simeq 10^{-4}~\text{Hz} \simeq 10^{-18}~\text{eV}$. This theoretical value is far from reached in practice due to all types of imperfections. Nonetheless, let us evaluate what could be expected in the case of a nanolaser, assuming this formula retains some physical relevance (at least in spirit). For a lasing nanoshell, as seen on Fig.~\ref{fg:polblw}-(a), we now have $\Delta\omega_\text{cav} \simeq 10^{-2}~\text{eV} \simeq 10^{6}~\text{MHz}$, while we may take $P_\text{out} \simeq 10^{-4}$~mW~\cite{Baranov:2013}, yielding $\Delta\omem \simeq 10^{12}~\text{Hz} \simeq 10^{-2}~\text{eV}$. This last value not only is immensely larger than the equivalent for a classical laser, but in fact, also lies in the same range as the spectral widths already exhibited in Figs.~\ref{fg:Grange} and \ref{fg:rho} (typically a few $10^{-2}$~eV). This would suggest that, should the classical scenario for homogeneously broadened transitions apply, the single surviving mode would barely sharpen; and therefore, the final emission width of the nanoshell could possibly not change much, if at all, in comparison to the maximal spectral width as we calculated it.

To definitely conclude on this point, the only way to calculate the actual final width of emission of the lasing nanoshell would be to compute the growth dynamics of the whole set of unstable frequencies in the spectrum, taken all together (not independently), fully accounting for their competitive effect on the population inversion $N$. This is a challenging task that we leave for future work. 

Nonetheless, we shall close this discussion by emphasizing that the (maximal) emission widths found in this work, which are in the range 5--10~nm (see Fig.~\ref{fg:Grange}), are as such already comparable to the typical values measured experimentally on actual nanolasers~\cite{Noginov:2009,Azzam:2020,Ma:2019,Ma:2021,Ellis:2024,Wu:2019,Yang:2017,Deeb:2017,Wang:2017,Lu:2014}.

\section{Conclusions}
To conclude, in this paper, we have unveiled a study of the properties of emission of a nanolaser in the nanoshell geometry, with gain in the core, and metal in the shell, both under and above the lasing threshold. For the first time, the free lasing regime was carefully studied, both in the dynamical transient regime and in the non-linear steady state, showing that strong frequency shifts effects (pull-out) shape up the spectrum of emission of the particle. These novel theoretical results add to the knowledge on one of the most promising geometries for nanolasers, in the hope to bring real-world applications within this thriving field one step closer.

\section{Methods} \label{sc:meth}
We here present some of the intermediate technical steps required to obtain the final set of governing equations in matrix form, as shown in eq~\ref{eq:qmastereq}--\ref{eq:Nmastereq}.

From the dipolar description of the nanoshell displayed in eq~\ref{ph1l1}--\ref{ps2}, one first needs to compute the amplitudes $p_{0,1,2,3}$ by enforcing continuity of the tangential electrical field and normal displacement at the boundaries $r=a$ and $r=\rho a$. This procedure yields the following expressions relating $p_{0,1,2,3}$ to $q_{0,1,2}$:
%(\epsilon_\text{b}+2\epsilon_\infty)(\epsilon_\infty+2\epsilon_s)+2\rho^3(\epsilon_\text{b}-\epsilon_\infty)(\epsilon_\infty-\epsilon_s)

\begin{align}
    p_3=& \frac{-\epsilon_{\infty} p_1 + 2 \epsilon_{\infty} \rho^3 p_2 - q_1 + 2 \rho^3 q_2 - \epsilon_e E_0 }{2 \epsilon_e}\label{eq:p3nn} \\
    p_2=&\frac{ \left( \epsilon_\text{b} - \epsilon_{\infty} \right) \left( p_3 - E_0 \right) + q_0 - q_1 + 2 q_2 }{- 2 \epsilon_{\infty} - \epsilon_\text{b} + \rho^3 \left( \epsilon_\text{b} - \epsilon_{\infty} \right) }\label{eq:p2nn}\\
    p_1=&p_3-\rho^3 p_2-E_0\label{eq:p1nn}\\
    p_0=&p_1+p_2.\label{eq:p0nn}
\end{align}
Detailed calculations to obtain these relations can be found in the Supplementary Information~\cite{SM}. Equations~\ref{eq:p3nn} to \ref{eq:p0nn} allow to calculate the time-dependent values of all electrical field components $p_i(t)$ from the knowledge of the polarization mode components $q_i(t)$. The latter are known from the resolution of the governing set of equations of the system, eq~\ref{eq:qmastereq}--\ref{eq:Nmastereq}.

We can now substitute equations~\ref{ph1l1}--\ref{ps2} and~\ref{eq:p3nn}--\ref{eq:p0nn} into the dynamical equations~\ref{eq:1ddn}--\ref{eq:3ccn}. Through this procedure, we produce a system of equations determining the time evolution of the mode amplitudes $q_{0,1,2}$ pertaining to the polarizations in the nanoparticle:
\begin{align}
      &\frac{dq_0}{dt}-\Omega_\text{g}q_0=\Gammag Np_0\,;\label{eq:q0nn}\\
      &\frac{dq_1}{dt}-\Omega_\text{m}q_1=\Gammam  p_1\,;\label{eq:q1nn}\\
      &\frac{dq_2}{dt}-\Omega_\text{m}q_2=\Gammam  p_2\,;\label{eq:q2nn}
\end{align}
and the equation for the evolution of the population inversion $N$:
\begin{equation}
      \frac{dN}{dt}+\frac{N-\widetilde{N}}{\tau_1}=\frac{1}{n\hbar}\text{Im}\left\{q_0p_0^*\right\}\,.\label{eq:Nn}
\end{equation}
In the set of equations~\ref{eq:q1nn}--\ref{eq:q2nn}, we have defined the shorthand notations:
\begin{align}
      &\Omega_\text{g}=i(\omega-\omg)-\frac{1}{\tau_2}\label{eq:oh}\\
      &\Gammag =-\frac{i\epsilon_0 G}{\widetilde{N}\tau_2}\label{eq:tg}\\
      &\Omega_\text{m}=\frac{\omega(\omega+2i\gamma)}{2(\gamma-i\omega)}\label{eq:op}\\
      &\Gammam  =\frac{\epsilon_0\omega_\text{p}^2}{2(\gamma-i\omega)}\label{eq:gp}.
\end{align}
Since relations~\ref{eq:p3nn}--\ref{eq:p0nn} make a linear system, we can write $p_0$, $p_1$, $p_2$ and $p_3$ as linear combinations of $q_0$, $q_1$, $q_2$ and $E_0$, namely:
\begin{align}
 p_0=& p_{00}q_0+p_{01}q_1+p_{02}q_2+p_{03}E_0,\label{eq:p0pl}\\
 p_1=& p_{10}q_0+p_{11}q_1+p_{12}q_2+p_{13}E_0,\label{eq:p1pl}\\
 p_2=& p_{20}q_0+p_{21}q_1+p_{22}q_2+p_{23}E_0,\\
 p_3=& p_{30}q_0+p_{31}q_1+p_{32}q_2+p_{33}E_0.\label{eq:p3pl}
\end{align}
The above coefficients $p_{ij}$ are real constants, whose analytical expressions involve combinations of the four parameters $\rho$, $\epsilon_\text{b}$, $\epsilon_\infty$ and $\epsilon_\text{e}$ only. (Full expressions are given in the Supplementary Information~\cite{SM}.)

We can then use these coefficients $p_{ij}$ to define the following matrix ${\bf A}(N)$:
\begin{equation}
{\bf A}(N)
=
\begin{bmatrix}
    \Gammag Np_{00}+\Omega_\text{g} & \Gammag Np_{01} & \Gammag Np_{02} \\
    \Gammam  p_{10} & \Omega_\text{m}+\Gammam p_{11} & \Gammam p_{12}\\
    \Gammam p_{20} & \Gammam p_{21} & \Omega_\text{m}+\Gammam p_{22}
\end{bmatrix} \label{eq:A}
\end{equation}
and the vector ${\bf b}(N,E_0)$:
\begin{equation}
 {\bf b}(N,E_0)= E_0 \bigl[\Gammag Np_{03},\Gammam p_{13}, \Gammam p_{23} \bigr]^\mathrm{T}.
\end{equation}
Collecting the mode amplitudes $q_i$ into a vector ${\bf q}$
\begin{equation}
    \label{eq:qdef}
    {\bf q}=\bigl[q_0, q_1, q_2\bigr]^\mathrm{T},
\end{equation}
the system of equations~\ref{eq:q1nn}--\ref{eq:Nn} can be rewritten in the following matrix form:
\begin{align}
 &\frac{d{\bf q}}{dt}={\bf A}(N)\cdot{\bf q}+{\bf b},\\
 &\frac{dN}{dt}+\frac{N-\widetilde{N}}{\tau_1}=\frac{1}{n\hbar}\text{Im}\left\{q_0p_0^*\right\},
\end{align}
This gives the governing set of equations for the nanoshell's dynamics, as shown in eq~\ref{eq:qmastereq}--\ref{eq:Nmastereq}.

We can see that the matrix ${\bf A} (N)$ encodes all the information about the nanoshell geometry of the system, via the coefficients $p_{ij}$. For other geometries like a homogeneous nanolaser sphere, or a core-shell one, the matrix $\bf{A}$ would admit different components from those of eq~\ref{eq:A}, but the global formalism of eq~\ref{eq:qmastereq}--\ref{eq:Nmastereq} will remain unchanged (as long as fields keep irrotational). Importantly, $\bf{A}$ also explicitly depends on the population inversion $N=N({\bf q},t)$, which in general is time-dependent, and, most importantly, depends non-linearly on the $q_i$ via eq~\ref{eq:Nmastereq} and relations~\ref{eq:p3nn}--\ref{eq:p0nn}. The $\bf A$ matrix also depends on the frequency $\omega$ and on the level of gain through the factor $\Gammag  \propto G$.

The vector ${\bf b}(N,E_0)$ depends on $N$ as well and on the excitation by the probe field $E_0$.

\section{Supporting information}
Complete model calculations and results; one additional figure pertaining to the ``Above threshold'' section (PDF).

\section{Author statements} 
 
All authors have accepted responsibility for the entire content of this manuscript and approved its submission.
 
Authors state no conflict of interest.

The datasets generated during and/or analyzed during the current study are available from the corresponding author on reasonable request.
\bibliographystyle{unsrt}
\bibliography{updated_2024}
\clearpage
\appendix
\section*{Supplementary Information}

\addcontentsline{toc}{section}{Supplementary Information}
\renewcommand{\theequation}{S.\arabic{equation}}
\setcounter{equation}{0}

\section{Extended Calculations}

We here present all the extended analytical calculations that support the results presented in the main text.

\subsection{Metal and Gain Medium Description}

We start by the description of the materials composing the nanoshell.

Please note that all fields written without a tilde in this section correspond to real-valued quantities (measurable in the physical world), while fields with a tilde represent the corresponding complex amplitudes.

{\bf 1. Metal}

We begin by describing how the field ${\bf E}_\text{m}({\bf r}, t)$, where $\bf r$ is the spatial coordinate with its origin at the particle center and $t$ is time, interacts with the electrons in the metallic nanoshell. This interaction is modeled using Drude's free-electron model:
\begin{equation}
\frac{d^2 \bf d}{d t^2} + 2 \gamma \frac{d \bf d}{dt} = \frac{e}{m_\text{e}} {\bf E}_\text{m}, \label{eq:Drude} \end{equation}
where $\bf d$ represents the displacement of the electron cloud from its equilibrium position, $m_\text{e}$ and $e$ are the electron mass and charge, respectively, and $\gamma$ is the ionic collision friction coefficient. We can then define the collective polarization produced by this displacement as:
\begin{equation}
    {\bf \Pi}_\text{m} = n_\text{e} e \bf d, \label{eq:Pimetal}
\end{equation}
here $n_\text{e}$ is the electron density in the metal. Substituting expression \ref{eq:Pimetal} into \ref{eq:Drude}, and considering that the plasma frequency is given by
\begin{equation}
    \omega^2_\text{p} = \frac{n_\text{e} e^2}{\epsilon_0 m_\text{e}},
\end{equation}
we can finally obtain the equation for the time evolution of ${\bf \Pi}_\text{m}$:
\begin{equation}
    \frac{d^2\mathbf{\Pi}_\text{m}}{dt^2}+2\gamma\frac{d\mathbf{\Pi}_\text{m}}{dt}=\epsilon_0\omega_\text{p}^2\mathbf{E}_\text{m}.
    \label{eqm2}
\end{equation}
$\bf \Pi_\text{m}$ represents the dynamic component of the polarization in the metal. The total polarization experienced by the metal also includes the passive contribution from the ionic lattice:
\begin{equation*}
    \mathbf{P}_\text{m}=\epsilon_0\chi_{\infty}\mathbf{E}_\text{m}+\mathbf{\Pi}_\text{m}.
\end{equation*}

Within the rotating wave approximation, the electric field and polarizations can be written in the following form:
\begin{align}
    \mathbf{E}_\text{m}(t) &= \frac{1}{2}\left[\mathbf{\tilde{E}}_\text{m}(t) e^{-i\omega t}+\mathbf{\tilde{E}}_\text{m}^*(t) e^{i\omega t}\right] \label{rw1m}\\
    \mathbf{\Pi}_\text{m}(t) &= \frac{1}{2}\left[\mathbf{\tilde{\Pi}}_\text{m}(t) e^{-i\omega t}+\mathbf{\tilde{\Pi}}_\text{m}^*(t) e^{i\omega t}\right] \\
    \mathbf{P}_\text{m}(t) &= \frac{1}{2}\left[\mathbf{\tilde{P}}_\text{m}(t) e^{-i\omega t}+\mathbf{\tilde{P}}_\text{m}^*(t) e^{i\omega t}\right],\label{rwem}
\end{align}

where $\mathbf{\tilde{E}}_\text{m}(t)$, $\mathbf{\tilde{\Pi}}_\text{m}(t)$, and $\mathbf{\tilde{P}}_\text{m}(t)$ represent a slow dependency on time (over times much slower than $1/\omega$).

If we now substitute expressions~\ref{rw1m}-\ref{rwem} into equation~\ref{eqm2} and average over fast time variations (times of order $1/\omega$), we can finally obtain the time evolution equation for the dynamic part of the polarization in the metal region:
\begin{equation}
    \frac{d{\bf \tilde{\Pi}}_\text{m}}{dt}-\frac{\omega(\omega+2i\gamma)}{2(\gamma-i\omega)}{\bf \tilde{\Pi}}_\text{m}=\frac{\epsilon_0\omega_\text{p}^2}{2(\gamma-i\omega)}{\bf \tilde{E}}_\text{m}, \label{eqm3}
\end{equation}
which is Eq.~\eqref{eq:3ccn} in the main article (where all tildas have been dropped by convention for readibility).

{\bf 2. Gain Medium}

The gain medium, made of emitters, can be modeled as a two-level system, where gain is achieved by introducing a phenomenological pump in addition to the typical thermal bath normally used to model purely absorbing elements. The two-level system is described through the optical Bloch equations in the density matrix formalism:
\begin{align}
    &\frac{d\rho_{21}}{dt}+\left(i\omega_{g}+\frac{1}{\tau_2}\right)\rho_{21}=-\frac{iN{\bm\mu}\cdot \mathbf{E}_\text{g}}{\hbar}
    \label{eq1}\\
   &\frac{dN}{dt}+\frac{N-\widetilde{N}}{\tau_1}=-\frac{2i(\rho_{21}-\rho_{12}){\bm\mu}\cdot \mathbf{E}_\text{g}}{\hbar}.
   \label{eq2}    
\end{align}
Here, the electric field of the gain medium, $\mathbf{E}_\text{g}$, interacts with a single gain element of dipole moment $\bm \mu$.
Also, $\rho_{ij}$ is the $i,j$ element of the density matrix. The time constants describing energy relaxation processes (spontaneous emission) and phase relaxation processes are, respectively, $\tilde{\tau}_1$ and $\tau_2$. We define the effective energy relaxation time $\tau_1$, which combines the effect of pumping and spontaneous emission on the population inversion $N$:
\begin{equation*}
    \tau_1=\frac{\tilde{\tau_1}}{W\tilde{\tau_1}+1}.
\end{equation*}
The transition frequency between levels 1 and 2 (of respective energies $E_1$ and $E_2$) is 
\begin{equation*}
    \omega_\text{g}=\frac{E_2-E_1}{\hbar}.
\end{equation*}
The quantity $N=\rho_{22}-\rho_{11}$ is the population inversion. When the gain element is subject to a phenomenological pump rate $W$, the corresponding equilibrium value of $N$ with the thermal reservoir is $N=\widetilde N$, given by
\begin{equation}\label{eq:tN}
    \widetilde{N}=\frac{W\tilde\tau_1-1}{W\tilde\tau_1+1}.
\end{equation}

The presence of $\widetilde{N}$ in Equation~\ref{eq2} means that, when the right-hand term of that equation is negligible, the population inversion is driven to $\tilde{N}$ in a time of the order of $\tau_1$. By choosing $\tilde{N}>0$ here, we are effectively modeling a pump that drives the active elements to their excited state.

In this framework, the polarization of the gain medium, as arising from the collective behavior of the population of gain elements, can be calculated as the following  integral: 
\begin{equation}
    \mathbf{P}_\text{g}=\epsilon_0\chi_\text{b}\mathbf{E}_\text{g}+\frac{n}{4\pi}\int_\Psi [\rho_{12}+\rho_{12}^*]{\bm\mu}d\Psi
    \label{P_h}
\end{equation}
where $\chi_\text{b}$ is the susceptibility of the dielectric host in which the gain elements are dispersed. The right side of expression~\ref{P_h} reflects the contribution of a population of gain elements with volume density $n$ and dipole moments $\bm\mu$ activated by the element of the density matrix $\rho_{12}$ and its conjugate, which account for the probability of transition. The distribution of dipoles is assumed to be randomly oriented, so that the expression is averaged over all solid angles $\Psi$ through the integral. Expression~\ref{P_h} shows that if the probability of transition were independent of the field in the gain region, the right term would just be averaged out. However, Equation~\ref{eq1} has a driving term on the right-hand side that favors the transition of the gain elements whose dipole moment is parallel to the electric field $\mathbf{E}_\text{g}$.

If we now define the active contribution to the polarization $\mathbf{\Pi}_\text{g}$ as:
\begin{equation}
    \mathbf{\Pi}_\text{g}=\frac{n}{4\pi}\int_\Psi [\rho_{12}+\rho_{12}^*]{\bm\mu}d\Psi,
\end{equation}
expression~\ref{P_h} can be rewritten as: 
\begin{equation}\label{eq:exp}
\mathbf{P}_\text{g}=\epsilon_0\chi_\text{b}\mathbf{E}_\text{g}+\mathbf{\Pi}_\text{g}.
\end{equation}
Also, considering that it is possible to demonstrate that
\begin{equation*}
    \int_\Psi({\bm\mu}\cdot\mathbf{E}_\text{g}){\bm\mu} d\Psi=\frac{4\pi}{3}\mu^2\mathbf{E}_\text{g},
\end{equation*}
one can rewrite the system of equations~\ref{eq1}-\ref{eq2} in terms of the time evolution of the dynamic part of the polarization in the gain medium:
\begin{align}
    &\frac{d\mathbf{\Pi}_\text{g}}{dt}+\left(i\omega_\text{g}+\frac{1}{\tau_2}\right)\mathbf{\Pi}_\text{g}=-\frac{2in\mu^2N}{3\hbar}\mathbf{E}_\text{g},
    \label{h1}\\
    &\frac{dN}{dt}+\frac{N-\widetilde{N}}{\tau_1}=\frac{i}{n\hbar}(\mathbf{\Pi}_\text{g}-\mathbf{\Pi}_\text{g}^*)\cdot \mathbf{E}_\text{g}.
    \label{h2}
\end{align}
Now, we use the rotating wave approximation again:
\begin{align*}
    \mathbf{E}_\text{g}(t)&=\frac{1}{2}\left[\mathbf{\tilde{E}}_\text{g}(t)e^{-i\omega t}+\mathbf{\tilde{E}}_\text{g}^*(t)e^{i\omega t}\right]\\
    \mathbf{\Pi}_\text{g}(t)&=\frac{1}{2}[\mathbf{\tilde{\Pi}}_\text{g}(t)e^{-i\omega t}+\mathbf{\tilde{\Pi}}_\text{g}(t)^*e^{i\omega t}]\\
    \mathbf{P}_\text{g}(t)&=\frac{1}{2}\left[\mathbf{\tilde{P}}_\text{g}(t)e^{-i\omega t}+\mathbf{\tilde{P}}_\text{g}^*(t)e^{i\omega t}\right],
\end{align*}
where $\mathbf{\tilde{E}}_h(t)$, $\mathbf{\tilde{\Pi}}_h(t)$ and $\mathbf{\tilde{P}}_h(t)$ represent again a slow dependency on time. When averaged over fast variations in time, \eqref{h1} and \eqref{h2} become:
\begin{align}
    &\frac{d\mathbf{\tilde{\Pi}}_\text{g}}{dt}-\left[i(\omega-\omega_\text{g})-\frac{1}{\tau_2}\right]\mathbf{\tilde{\Pi}}_\text{g}=-\frac{in\mu^2N}{3\hbar}\mathbf{\tilde{E}}_\text{g},
    \label{h11}\\
    &\frac{dN}{dt}+\frac{N-\widetilde{N}}{\tau_1}=-\frac{i}{2n\hbar}(\mathbf{\tilde{\Pi}}_\text{g}\cdot\mathbf{\tilde{E}}_\text{g}^*-\mathbf{\tilde{\Pi}}_\text{g}^*\cdot\mathbf{\tilde{E}}_\text{g}) .
    \label{h22}
\end{align}

By defining the parameter $G$, which gives a measure of the level of gain brought into the system by the gain medium elements under pumping: 
\begin{equation}
    G= \frac{\tau_2\mu^2}{3\hbar\epsilon_0}n\widetilde{N},
\end{equation}
one can rewrite the system of equations~\ref{h11}-\ref{h22} as:
\begin{align}
      &\frac{d{\bf \tilde{\Pi}}_\text{g}}{dt}-\left[i(\omega-\omega_\text{g})-\frac{1}{\tau_2}\right]{\bf \tilde{\Pi}}_\text{g}=-\frac{i\epsilon_0 G}{\tau_2}\frac{N}{\widetilde{N}}{\bf \tilde{E}}_\text{g},\label{eq:s:H}\\
      &\frac{dN}{dt}+\frac{N-\widetilde{N}}{\tau_1}=-\frac{i}{2n\hbar}({\bf\tilde{\Pi}}_\text{g}\cdot{\bf \tilde{E}}^*_\text{g}-{\bf\tilde{\Pi}}_\text{g}^*\cdot{\bf \tilde{E}}_\text{g}).\label{eq:s:N}
\end{align}

This system of equations governs the time evolution of the gain-enriched medium for different amounts of the gain quantity $G$. These equations are the same as Eqs.~\eqref{eq:1ddn}--\eqref{eq:2ddn} in the main article (where it is reminded that all tildas were dropped out of notational convenience).

\subsection{Steady-State Permittivities}\label{s:A}

From this point onwards, tildas will be meant implicitly for all fields and polarizations vectors and shall be removed, i.e., we are now exclusively dealing with the slowly-evolving, complex amplitudes introduced in the rotating-wave approximation.

When and if equation~\ref{eqm3} and system~\ref{eq:s:H}-\ref{eq:s:N} reach a steady state, one can calculate the permittivities $\epsilon_\text{g}$ and $\epsilon_\text{m}$ for the gain medium and the metal.

Starting with the metal permittivity, let us first consider the steady-state solution of equation \ref{eqm3}:
\[
    -\frac{\omega(\omega+2i\gamma)}{2(\gamma-i\omega)}{\bf \Pi}_\text{m} = \frac{\epsilon_0 \omega_\text{p}^2}{2(\gamma-i\omega)}{\bf E}_\text{m}, 
\]
from which one can calculate:
\[{\bf \Pi}_\text{m} = - \frac{\epsilon_0 \omega_\text{p}^2}{\omega (\omega + 2i \gamma)}{\bf E}_\text{m}.
\]
Replacing the previous result in equation \ref{pm}, one gets:
\begin{equation*}
    {\bf P}_\text{m} = \epsilon_0 \left[ \chi_\infty - \frac{\omega_\text{p}^2}{\omega (\omega + 2i \gamma)} \right] {\bf E}_\text{m}.
\end{equation*}
Thus, the electric displacement is:
\begin{align*}
    {\bf D}_\text{m} &= \epsilon_0 {\bf E}_\text{m} + {\bf P}_\text{m} \\
    {\bf D}_\text{m} &= \epsilon_0 \left[ 1 + \chi_\infty - \frac{\omega_\text{p}^2}{\omega (\omega + 2i \gamma)} \right] {\bf E}_\text{m},
\end{align*}
meaning that the metal steady state permittivity is:
\begin{equation}
 \epsilon_\text{m} = \epsilon_\infty - \frac{\epsilon_0 \omega_\text{p}^2}{\omega(\omega+2i\gamma)},\label{eq:s:dru}
\end{equation}
where $\epsilon_\infty = \epsilon_0 ( 1 + \chi_\infty)$. Expression~\ref{eq:s:dru} can be recognized as the Drude formula for metal permittivity, which appears as Eq.~\eqref{eq:dru} in the main article.

Let us now switch to the gain medium. The steady-state solution of equation \ref{eq:s:H} is:
\[
    - \left[ i \left( \omega - \omega_\text{g} \right) - \dfrac{1}{\tau_2} \right] {\bf \Pi}_\text{g} = - \frac{i \epsilon_0 G N}{\tilde{N} \tau_2} {\bf E}_\text{g},
\]

from which one can calculate
\begin{align}
    {\bf \Pi}_\text{g} &= \frac{i \epsilon_0 G N}{\tilde{N} \tau_2} \frac{1}{i \left( \omega - \omega_\text{g} \right) - \dfrac{1}{\tau_2}} {\bf E}_\text{g} \\
     &= \frac{\epsilon_0 G N \Delta}{\tilde{N}} \frac{1}{2 \left( \omega - \omega_\text{g} \right) + i \Delta} {\bf E}_\text{g},  \label{eq:Pih}
\end{align}
where we have defined the gain linewidth $\Delta=2/\tau_2$.

Therefore, replacing this expression into the expression for the electric displacement in the gain medium, we get
\begin{align*}
    {\bf D}_\text{g} &= \epsilon_0 {\bf E}_\text{g} + {\bf P}_\text{g} \\
    &= \epsilon_\text{b} {\bf E}_\text{g} + {\bf \Pi}_\text{g} \\
    &= \left[ \epsilon_\text{b} + \frac{\epsilon_0 G N \Delta}{\tilde{N}} \frac{1}{2 \left( \omega - \omega_\text{g} \right) + i \Delta} \right] {\bf E}_\text{g},
\end{align*}
where we define $\epsilon_\text{b} = \epsilon_0(1 + \chi_\text{b})$, and the permittivity of the gain medium is:
\begin{equation}
    \epsilon_\text{g} = \epsilon_\text{b} + \frac{\epsilon_0 G \Delta}{2 \left( \omega - \omega_\text{g} \right) + i \Delta}\,\frac{N}{\tilde{N}}. \label{eq:epsgN}
\end{equation}

To obtain the expression for $N$, we calculate the steady state solution of equation \ref{eq:s:N}, which is:
\begin{equation}
    N = \tilde{N} - \frac{i \tau_1}{2 n \hbar} \left( {\bf \Pi}_\text{g} \cdot {\bf E}_\text{g}^* - {\bf \Pi}_\text{g}^* \cdot {\bf E}_\text{g} \right). \label{eq:eNe}
\end{equation}
By using equation \ref{eq:Pih}, we can calculate the right side of equation \ref{eq:eNe}, and obtain $N$:
\begin{align*}
    &N = \tilde{N} - \dfrac{\epsilon_0G \tau_1 \Delta^2}{ n \hbar \tilde{N}} N \frac{1}{ \Delta^2 + 4 \left( \omega - \omega_\text{g} \right)^2 } |{\bf E}_\text{g}|^2 \\
    &N \left[ \frac{\Delta^2 + 4 \left( \omega - \omega_\text{g} \right)^2 + \dfrac{\epsilon_0G \tau_1 \Delta^2}{ n \hbar \tilde{N}}|{\bf E}_\text{g}|^2}{\Delta^2 + 4 \left( \omega - \omega_\text{g} \right)^2} \right] = \tilde{N} \\
    &N = \tilde{N} \frac{\Delta^2 + 4 \left( \omega - \omega_\text{g} \right)^2}{\Delta^2 + 4 \left( \omega - \omega_\text{g} \right)^2 + \dfrac{\epsilon_0G \tau_1 \Delta^2}{ n \hbar \tilde{N}}|{\bf E}_\text{g}|^2}. 
\end{align*}
By introducing $E_\text{sat}=\sqrt{n \hbar \tilde{N}/(\epsilon_0 G\tau_1)}$, which can be rewritten as $E_\text{sat}=\hbar/\mu\sqrt{3/(\tau_1\tau_2)}$
\begin{equation}
    N = \tilde{N} \frac{4 \left( \omega - \omega_\text{g} \right)^2+\Delta^2}{4 \left( \omega - \omega_\text{g} \right)^2 + \Delta^2\left(1+\dfrac{|{\bf E}_\text{g}|^2}{{E_\text{sat}}^2}\right)}. \label{eq:Nagain}
\end{equation}
By replacing \ref{eq:Nagain} in equation \ref{eq:Pih}, we obtain:
\begin{align*}
    {\bf \Pi}_\text{g} &= \frac{\epsilon_0 G \Delta}{2 \left( \omega - \omega_\text{g} \right) + i \Delta} \frac{4 \left( \omega - \omega_\text{g} \right)^2+\Delta^2}{4 \left( \omega - \omega_\text{g} \right)^2 + \Delta^2\left(1+\dfrac{|{\bf E}_\text{g}|^2}{{E_\text{sat}}^2}\right)} {\bf E}_\text{g} \\
    &= \epsilon_0 G \Delta \frac{2 \left( \omega - \omega_\text{g} \right) - i \Delta}{4 \left( \omega - \omega_\text{g} \right)^2 + \Delta^2\left(1+\dfrac{|{\bf E}_\text{g}|^2}{{E_\text{sat}}^2}\right)} {\bf E}_\text{g}.
\end{align*}

Now, we are able to calculate the electric displacement in the gain medium:
\begin{equation*}
    {\bf D}_\text{g} = \epsilon_0 {\bf E}_\text{g} + {\bf P}_\text{g} = \epsilon_0 \left[ 1 + \chi_\text{b} + \frac{\left[ 2 \left( \omega - \omega_\text{g} \right) - i \Delta \right] G \Delta}{4 \left( \omega - \omega_\text{g} \right)^2 + \Delta^2\left(1+\dfrac{|{\bf E}_\text{g}|^2}{{E_\text{sat}}^2}\right)} \right] {\bf E}_\text{g},
\end{equation*}
from which we determine the permittivity, where $\epsilon_\text{b} = \epsilon_0(1 + \chi_\text{b})$:
\begin{equation}\label{eq:supsat}
    \epsilon_\text{g} = \epsilon_\text{b} + \frac{\epsilon_0\left[ 2 \left( \omega - \omega_\text{g} \right) - i \Delta \right]  G \Delta}{4 \left( \omega - \omega_\text{g} \right)^2 + \Delta^2\left(1+\dfrac{|{\bf E}_\text{g}|^2}{{E_\text{sat}}^2}\right)}.
\end{equation}

which is the expression for permittivity given in Eq.~\eqref{eq:slor} of the main text.

Equation \ref{eq:supsat} is the permittivity of the gain media in the saturated case, i.e. when $N \neq \widetilde{N}$. On the other hand, in the ``small-signal" regime, i.e. when $N = \tilde{N}$, we recover the linear Lorentzian permittivity:
\begin{equation*}
    \epsilon_\text{g}= \epsilon_\text{b}+\frac{\epsilon_0 G\Delta}{2(\omega-\omega_\text{g})+i\Delta}.
\end{equation*}

\subsection{Boundary Conditions}
The use of boundary conditions allows us to determine the coefficients of the Legendre polynomials present in the potentials defining the polarization and the electric field in the different regions of the system.

To proceed with the boundary conditions, we first calculate the radial and polar spherical coordinates components of the electric fields and polarizations, as derived from the potentials written down in equations~\eqref{ph1l1} to \eqref{ps2} from the main article:
\begin{align*}
    &E_\text{g}^r = - \frac{\partial \phi_1}{\partial r} = - p_0 \cos \theta \\
    &E_\text{g}^{\theta} = - \frac{1}{r} \frac{\partial \phi_1}{\partial \theta} = p_0 \sin \theta \\
    &\Pi_\text{g}^r = - \frac{\partial \psi_1}{\partial r} = - q_0 \cos \theta \\
    &\Pi_\text{g}^{\theta} = - \frac{1}{r} \frac{\partial \psi_1}{\partial \theta} = q_0 \sin \theta
\end{align*}
\begin{align*}
    &E_\text{m}^r = - \frac{\partial \phi_2}{\partial r} = - p_1 \cos \theta + 2 a^3 \rho^3 p_2 \frac{\cos \theta}{r^3} \\
    &E_\text{m}^{\theta} = - \frac{1}{r} \frac{\partial \phi_2}{\partial \theta} = p_1 \sin \theta + a^3 \rho^3 p_2 \frac{\sin \theta}{r^3} \\
    &\Pi_\text{m}^r = - \frac{\partial \psi_2}{\partial r} = - q_1 \cos \theta + 2 a^3 \rho^3 q_2 \frac{\cos \theta}{r^3} \\
    &\Pi_\text{m}^{\theta} = - \frac{1}{r} \frac{\partial \psi_2}{\partial \theta} = q_1 \sin \theta + a^3 \rho^3 q_2 \frac{\sin \theta}{r^3}
\end{align*}
\begin{align*}
    &E_\text{e}^r = - \frac{\partial \phi_3}{\partial r} = E_0 \cos \theta + 2 a^3 p_3 \frac{\cos \theta}{r^3} \\
    &E_\text{e}^{\theta} = - \frac{1}{r} \frac{\partial \phi_3}{\partial \theta} = - E_0 \sin \theta + a^3 p_3 \frac{\sin \theta}{r^3}.
\end{align*}

\newpage
{\bf 1. Metal Outer boundary $\bm{r = a}$:}

$\bullet$ Radial continuity:
\begin{align*}
    &D_\text{m}^r \vert_{r = a} = D_\text{e}^r \vert_{r = a} \\
    &\left( \epsilon_0 E_\text{m}^r + P_\text{m}^r \right) \vert_{r = a} = \left( \epsilon_0 E_\text{e}^r + P_\text{e}^r \right) \vert_{r = a} \\
    &\epsilon_{\infty} E_\text{m}^r \vert_{r = a} + \Pi_\text{m}^r \vert_{r = a} = \epsilon_\text{e} E_\text{e}^r \vert_{r = a} \\
    &\epsilon_{\infty} \left( -p_1 \cos \theta + 2 \rho^3 p_2 \cos \theta \right) - q_1 \cos \theta + 2 \rho^3 q_2 \cos \theta = \epsilon_\text{e} \left( E_0 \cos \theta + 2 p_3 \cos \theta \right) \\
    &- \epsilon_{\infty} p_1 + 2 \epsilon_{\infty} \rho^3 p_2 - q_1 + 2 \rho^3 q_2 = \epsilon_\text{e} E_0 + 2 \epsilon_\text{e} p_3 
\end{align*}
$\bullet$ Tangential continuity:
\begin{align*}
    &E_\text{m}^{\theta} \vert_{r = a} = E_\text{e}^{\theta} \vert_{r = a} \\
    &p_1 \sin \theta + \rho^3 \sin \theta p_2 = - E_0 \sin \theta + \sin \theta p_3 \\
    &p_1 + \rho^3 p_2 = - E_0 + p_3 
\end{align*}

{\bf 2. Gain-metal boundary at $\bm{r = \rho a}$:}
$\bullet$ Radial continuity:
\begin{align*}
    &D_\text{m}^r \vert_{r = \rho a} = D_\text{g}^r \vert_{r = \rho a} \\
    &\left( \epsilon_0 E_\text{m}^r + P_\text{m}^r \right) \vert_{r = \rho a} = \left( \epsilon_0 E_\text{g}^r + P_\text{g}^r \right) \vert_{r = \rho a} \\
    &\epsilon_{\infty} E_\text{m}^r \vert_{r = \rho a} + \Pi_\text{m}^r \vert_{r = \rho a} = \epsilon_\text{b} E_\text{g}^r \vert_{r = \rho a} + \Pi_\text{g}^r \vert_{r = \rho a} \\
    &\epsilon_{\infty} \left( -p_1 \cos \theta + 2 \rho^3 a^3 p_2 \frac{\cos \theta}{\rho^3 a^3} \right) - q_1 \cos \theta + 2 \rho^3 a^3 q_2 \frac{\cos \theta}{\rho^3 a^3} = - \epsilon_\text{b} p_0 \cos \theta - q_0 \cos \theta \\
    &- \epsilon_{\infty} p_1 + 2 \epsilon_{\infty} p_2 - q_1 + 2 q_2 = - \epsilon_\text{b} p_0 - q_0  
\end{align*}

$\bullet$ Tangential continuity:
\begin{align*}
    &E_\text{m}^{\theta} \vert_{r = \rho a} = E_\text{g}^{\theta} \vert_{r = \rho a} \\
    &p_1 \sin \theta + \frac{\rho^3 a^3 \sin \theta}{\rho^3 a^3} p_2 = p_0 \sin \theta \\
    &p_1 + p_2 = p_0  \label{eq:tangentialrhoa}
\end{align*}

Therefore, from the boundary conditions, we obtain:
\begin{align}
    p_3=& \frac{-\epsilon_{\infty} p_1 + 2 \epsilon_{\infty} \rho^3 p_2 - q_1 + 2 \rho^3 q_2 - \epsilon_\text{e} E_0 }{2 \epsilon_\text{e}} \\
    p_2=&\frac{ \left( \epsilon_\text{b} - \epsilon_{\infty} \right) \left( p_3 - E_0 \right) + q_0 - q_1 + 2 q_2 }{- 2 \epsilon_{\infty} - \epsilon_\text{b} + \rho^3 \left( \epsilon_\text{b} - \epsilon_{\infty} \right) } \\
    p_1=&p_3-\rho^3 p_2-E_0 \\
    p_0=&p_1+p_2.
\end{align}

\subsection{Steady-State Polarizability $\pmb \alpha$}\label{s:B}

Let us now prove that Eq.~\eqref{eq:p3alpha} in the main article holds, with the classical expression for the polarizability $\alpha$ as written in Eq.~\eqref{eq:polarizability} :
\begin{equation}\label{pol}
 \frac{\alpha}{4\pi a^3}=\frac{(\epsilon_\text{m}-\epsilon_\text{e})(\epsilon_\text{g}+2\epsilon_\text{m})+\rho^3(\epsilon_\text{g}-\epsilon_\text{m})(\epsilon_\text{e}+2\epsilon_\text{m})}{(\epsilon_\text{g}+2\epsilon_\text{m})(\epsilon_\text{m}+2\epsilon_\text{e})+2\rho^3(\epsilon_\text{g}-\epsilon_\text{m})(\epsilon_\text{m}-\epsilon_\text{e})}
\end{equation}

To demonstrate this statement, we begin with calculating the steady-state solutions of equations~\eqref{eq:q0nn} to \eqref{eq:q2nn} of the main article:
\begin{align}
    &q_0 = - \frac{\Gamma_\text{g}N}{\Omega_\text{g}} p_0 \\
    &q_1 = - \frac{\Gamma_\text{m}}{\Omega_\text{m}} p_1 \\
    &q_2 = - \frac{\Gamma_\text{m}}{\Omega_\text{m}} p_2.
\end{align}
From equation~\eqref{eq:oh}~of the article, we can deduce that
\begin{equation*}
    - 2 i \Omega_\text{g} = 2 \left( \omega - \omega_\text{g} \right) + i \Delta.
\end{equation*}
Recall now the expression for the permittivity of the gain medium in equation \ref{eq:epsgN}:
\begin{equation}
    \epsilon_\text{g} = \epsilon_\text{b} + \frac{\epsilon_0 G N \Delta}{\tilde{N}} \frac{1}{2 \left( \omega - \omega_\text{g} \right) + i \Delta}. \label{eq:epsgN1}
\end{equation}
Replacing the expression for $- 2 i \Omega_\text{g}$ in equation \ref{eq:epsgN1}, we obtain that
\begin{equation}
    \epsilon_\text{g} = \epsilon_\text{b} - \frac{\epsilon_0 G N \Delta}{\tilde{N}} \frac{1}{2 i \Omega_\text{g}}. \label{eq:GDelta}
\end{equation}
Also, according to equation~\eqref{eq:tg} in the article:
\begin{equation*}
    G = \frac{2 i \tilde{N}}{\epsilon_0 \Delta} \Gamma_\text{g},
\end{equation*}
\begin{comment}
    Replacing $- 2 i \Omega_\text{g}$ in equation \ref{eq:lor}, we obtain that
\begin{align*}
    &\epsilon_\text{g} = \epsilon_\text{b} - \frac{G \Delta}{2 i \Omega_\text{g}} \\
    &G \Delta = 2 i \Omega_\text{g} \left( \epsilon_\text{b} - \epsilon_\text{g} \right)  \label{eq:GDelta}.
\end{align*}
\end{comment}
thus, equation~\eqref{eq:epsgN} becomes
\begin{equation*}
    \epsilon_\text{g} - \epsilon_\text{b} = - \frac{N \Gamma_\text{g}}{\Omega_\text{g}}.
\end{equation*}
\begin{comment}
    When neglecting the right side of equation \ref{eq:Nn}, its steady-state solution is $N = \tilde{N}$. Therefore, from equation \ref{eq:tg} we obtain that
\begin{equation}
    \Gamma_\text{g} N = - \frac{i \epsilon_0 G \Delta}{2} \label{eq:GN}.
\end{equation}
Replacing equation \ref{eq:GDelta} into equation \ref{eq:GN}, we get
\begin{align*}
    &\Gamma_\text{g}N = \epsilon_0 \Omega_\text{g} \left( \epsilon_\text{b} - \epsilon_\text{g} \right) \\
    &\frac{\Gamma_\text{g}N}{\Omega_\text{g}} = \epsilon_\text{b} - \epsilon_\text{g}.
\end{align*}
\end{comment}
On the other hand, from equations~\eqref{eq:op} and \eqref{eq:gp}, we obtain that
\begin{align*}
    \frac{\Gamma_\text{m}}{\Omega_\text{m}} &= \frac{\epsilon_0 \omega^2_\text{p}}{\omega \left( \omega + 2i \gamma \right)} \\
    &= \epsilon_{\infty} - \epsilon_\text{m}.
\end{align*}
Consequently, $q_0$, $q_1$, and $q_2$ can be written as
\begin{align}
\begin{split}
    &q_0 = \left( \epsilon_\text{g} - \epsilon_\text{b} \right) p_0 \\
    &q_1 = \left( \epsilon_\text{m} - \epsilon_{\infty} \right) p_1 \\
    &q_2 = \left( \epsilon_\text{m} - \epsilon_{\infty} \right) p_2.
\end{split} \label{eq:setq}
\end{align}

We now simplify equation~\eqref{eq:p3nn} of the main article by substituting the set of equations \ref{eq:setq} into it:
\begin{align}
    p_3=& \frac{-\epsilon_{\infty} p_1 + 2 \epsilon_{\infty} \rho^3 p_2 - \left( \epsilon_\text{m} - \epsilon_{\infty} \right) p_1 + 2 \rho^3 \left( \epsilon_\text{m} - \epsilon_{\infty} \right) p_2 - \epsilon_\text{e} E_0 }{2 \epsilon_\text{e}} \\
    =& \frac{- \epsilon_\text{m} p_1 + 2 \rho^3 \epsilon_\text{m} p_2 - \epsilon_\text{e} E_0}{2 \epsilon_\text{e}}.  \label{eq:p3simpl}
\end{align}
Next, we simplify equation~\eqref{eq:p2nn} by using equation \eqref{eq:p1nn} and substituting with the set \ref{eq:setq} again:
\begin{align*}
    p_2=&\frac{ \left( \epsilon_\text{b} - \epsilon_{\infty} \right) \left( p_3 - E_0 \right) + \left( \epsilon_\text{g} - \epsilon_\text{b} \right) p_0 - \left( \epsilon_\text{m} - \epsilon_{\infty} \right) p_1 + 2 \left( \epsilon_\text{m} - \epsilon_{\infty} \right) p_2 }{- 2 \epsilon_{\infty} - \epsilon_\text{b} + \rho^3 \left( \epsilon_\text{b} - \epsilon_{\infty} \right) }\\
    =&\frac{ \left( \epsilon_\text{g} - \epsilon_\text{b} - \epsilon_\text{m} + \epsilon_{\infty} \right) \left( p_3 - \rho^3 p_2 - E_0 \right) + \left( \epsilon_\text{g} - \epsilon_\text{b} + 2 \epsilon_\text{m} - 2 \epsilon_{\infty} \right) p_2 }{- 2 \epsilon_{\infty} - \epsilon_\text{b} + \rho^3 \left( \epsilon_\text{b} - \epsilon_{\infty} \right) }\\
    =& \frac{\left( \epsilon_\text{m} - \epsilon_\text{g} \right) \left( p_3 - E_0 \right) }{ \epsilon_\text{g} + 2 \epsilon_\text{m} + \rho^3 \left( \epsilon_\text{m} - \epsilon_\text{g} \right) }.
\end{align*}
We replace this result in equation \ref{eq:p1nn} to obtain $p_1$ in terms of $p_3$ and $E_0$:
\begin{align*}
    p_1 &= - E_0 - \frac{ \rho^3 \left( \epsilon_\text{m} - \epsilon_\text{g} \right) \left( p_3 - E_0 \right) }{ \epsilon_\text{g} + 2 \epsilon_\text{m} + \rho^3 \left( \epsilon_\text{m} - \epsilon_\text{g} \right) } + p_3 \\
    &= \frac{\left( \epsilon_\text{g} + 2 \epsilon_\text{m} \right) \left( p_3 - E_0 \right) }{ \epsilon_\text{g} + 2 \epsilon_\text{m} + \rho^3 \left( \epsilon_\text{m} - \epsilon_\text{g} \right) }.
\end{align*}
By using these expressions of $p_2$ and $p_1$ in equation \ref{eq:p3simpl}, we obtain:
\begin{equation*}
    p_3 = - \frac{\epsilon_\text{m}}{2 \epsilon_\text{e}} \frac{\left( \epsilon_\text{g} + 2 \epsilon_\text{m} \right) \left( p_3 - E_0 \right) }{ \epsilon_\text{g} + 2 \epsilon_\text{m} + \rho^3 \left( \epsilon_\text{m} - \epsilon_\text{g} \right) } + \frac{\rho^3 \epsilon_\text{m}}{\epsilon_\text{e}} \frac{\left( \epsilon_\text{m} - \epsilon_\text{g} \right) \left( p_3 - E_0 \right) }{ \epsilon_\text{g} + 2 \epsilon_\text{m} + \rho^3 \left( \epsilon_\text{m} - \epsilon_\text{g} \right) } - \frac{E_0}{2},
 \end{equation*}
which after rearrangement, gives the proportionality relation between $p_3$ and $E_0$:
\begin{equation}
    p_3 = \frac{ \left( \epsilon_\text{g} + 2 \epsilon_\text{m} \right) \left( \epsilon_\text{m} - \epsilon_\text{e} \right) + \rho^3 \left( \epsilon_\text{g} - \epsilon_\text{m} \right) \left( \epsilon_\text{e} + 2 \epsilon_\text{m} \right) }{ \left( \epsilon_\text{g} + 2 \epsilon_\text{m} \right) \left( 2 \epsilon_\text{e} + \epsilon_\text{m} \right) + 2 \rho^3 \left( \epsilon_\text{g} - \epsilon_\text{m} \right) \left( \epsilon_\text{m} - \epsilon_\text{e} \right) } E_0.
\end{equation}
Since by definition of the polarizability [see Eqs.~\eqref{eq:dipmoment} and \eqref{eq:dipolalpha} of the article], we have $p_3= \alpha E_0/(4\pi a^3)$, we deduce that $\alpha$ has indeed the form of equation~\ref{pol}, or Eq.~\eqref{eq:polarizability} in the main text.

\subsection{The Geometry Matrix}

To express the matrix system of equations as stated in~\eqref{eq:A} of the article, let us recall expressions~\eqref{eq:p3nn}--\eqref{eq:p0nn}:
\begin{align*}
    p_3=& \frac{-\epsilon_{\infty} p_1 + 2 \epsilon_{\infty} \rho^3 p_2 - q_1 + 2 \rho^3 q_2 - \epsilon_\text{e} E_0 }{2 \epsilon_\text{e}} \\
    p_2=&\frac{ \left( \epsilon_\text{b} - \epsilon_{\infty} \right) \left( p_3 - E_0 \right) + q_0 - q_1 + 2 q_2 }{- 2 \epsilon_{\infty} - \epsilon_\text{b} + \rho^3 \left( \epsilon_\text{b} - \epsilon_{\infty} \right) }; \\
    p_1=&p_3-\rho^3 p_2-E_0; \\
    p_0=&p_1+p_2.
\end{align*}

It is now necessary to write all of these relations as linear  functions of the main variables $q_0$, $q_1$, $q_2$, as well as $E_0$.

By replacing equations~\eqref{eq:p1nn} and \eqref{eq:p2nn} into equation~\eqref{eq:p3nn}, we get:
\begin{align}
     p_3 =& \frac{\epsilon_{\infty}}{2 \epsilon_\text{e}} \left( E_0 + \rho^3 p_2 - p_3 \right) + \frac{2 \rho^3 \epsilon_{\infty}}{2 \epsilon_\text{e}} \frac{ \left( \epsilon_\text{b} - \epsilon_{\infty} \right) \left( p_3 - E_0 \right) + q_0 - q_1 + 2 q_2 }{- 2 \epsilon_{\infty} - \epsilon_\text{b} + \rho^3 \left( \epsilon_\text{b} - \epsilon_{\infty} \right) } \\
     &+ \frac{- q_1 + 2 \rho^3 q_2 - \epsilon_\text{e} E_0}{2 \epsilon_\text{e}} \\
    =&  -\frac{3\rho^3\epsilon_\infty}{D}q_0-\frac{(1-\rho^3)(\epsilon_\text{b}+2\epsilon_\infty)}{D}q_1+\frac{2\rho^3(1-\rho^3)(\epsilon_\text{b}-\epsilon_\infty)}{D}q_2 \\\
      &+\frac{(\epsilon_\infty-\epsilon_\text{e})(\epsilon_\text{b}+2\epsilon_\infty)+\rho^3(\epsilon_\text{b}-\epsilon_\infty)(\epsilon_\text{e}+2\epsilon_\infty)}{D}E_0,  \label{eq:ptres}
\end{align}
where, in order to have more compact formulas, we define:
\begin{equation*}
 D=(\epsilon_\infty+2\epsilon_\text{e})(\epsilon_\text{b}+2\epsilon_\infty)+2\rho^3(\epsilon_\text{b}-\epsilon_\infty)(\epsilon_\infty-\epsilon_\text{e}).
\end{equation*}

We then replace equation \ref{eq:ptres} into equation~\eqref{eq:p2nn}, and obtain:
\begin{equation}
    p_2=-\frac{\epsilon_\infty+2\epsilon_\text{e}}{D}q_0+\frac{\epsilon_\text{b}+2\epsilon_\text{e}}{D}q_1-\frac{2[(\epsilon_\infty+2\epsilon_2)+\rho^3(\epsilon_\text{b}-\epsilon_\infty)]}{D}q_2+\frac{3\epsilon_2(\epsilon_\text{b}-\epsilon_\infty)}{D}E_0. \label{eq:pdos}
\end{equation}

We can now calculate $p_1$ by replacing \ref{eq:ptres} and \ref{eq:pdos} into~\eqref{eq:p1nn}:
\begin{align}
    p_1 =\frac{2\rho^3(\epsilon_\text{e}-\epsilon_\infty)}{D}q_0-&\frac{\rho^3(\epsilon_\text{b}+2\epsilon_\text{e})+(1-\rho^3)(\epsilon_\text{b}+2\epsilon_\infty)}{D}q_1+\nonumber\\+&\frac{2\rho^3(\epsilon_\text{b}+2\epsilon_\text{e})}{D}q_2 
    -\frac{3\epsilon_\text{e}(\epsilon_\text{b}+2\epsilon_\infty)}{D}E_0.
    \label{eq:puno}
\end{align}

Finally, we calculate $p_0$:
\begin{align}
    p_0=-\frac{(\epsilon_\infty+2\epsilon_\text{e})+2\rho^3(\epsilon_\infty-\epsilon_\text{e})}{D}q_0-&\frac{2(1-\rho^3)(\epsilon_\infty-\epsilon_\text{e})}{D}q_1+\nonumber\\-&\frac{2(1-\rho^3)(\epsilon_\infty+2\epsilon_\text{e})}{D}q_2-\frac{9\epsilon_\text{e}\epsilon_\infty}{D}E_0.
    \label{eq:pzero}
\end{align}

We can thus rewrite the obtained expressions~\ref{eq:ptres}--\ref{eq:pzero} for $p_3$, $p_2$, $p_1$, and $p_0$ in the following form:
\begin{align*}
 p_0=& p_{00}q_0+p_{01}q_1+p_{02}q_2+p_{03}E_0\\
 p_1=& p_{10}q_0+p_{11}q_1+p_{12}q_2+p_{13}E_0\\
 p_2=& p_{20}q_0+p_{21}q_1+p_{22}q_2+p_{23}E_0\\
 p_3=& p_{30}q_0+p_{31}q_1+p_{32}q_2+p_{33}E_0.
\end{align*}

where the $p_{ij}$ coefficients have been  defined as:
\begin{align*}
 &p_{00}=-\frac{\epsilon_\infty+2\epsilon_\text{e}+2\rho^3(\epsilon_\infty-\epsilon_\text{e})}{D}\\
 &p_{01}=-\frac{2(1-\rho^3)(\epsilon_\infty-\epsilon_\text{e})}{D}\\
 &p_{02}=-\frac{2(1-\rho^3)(\epsilon_\infty+2\epsilon_\text{e})}{D}\\
 &p_{03}=-\frac{9\epsilon_\text{e}\epsilon_\infty}{D}\\
 &p_{10}=\frac{2\rho^3(\epsilon_\text{e}-\epsilon_\infty)}{D}\\
 &p_{11}=-\frac{\rho^3(\epsilon_\text{b}+2\epsilon_\text{e})+(1-\rho^3)(\epsilon_\text{b}+2\epsilon_\infty)}{D}\\
 &p_{12}=\frac{2\rho^3(\epsilon_\text{b}+2\epsilon_\text{e})}{D}\\
 &p_{13}=-\frac{3\epsilon_\text{e}(\epsilon_\text{b}+2\epsilon_\infty)}{D}\\
 &p_{20}=-\frac{\epsilon_\infty+2\epsilon_\text{e}}{D}\\
 &p_{21}=\frac{\epsilon_\text{b}+2\epsilon_\text{e}}{D}\\
 &p_{22}=-\frac{2[\epsilon_\infty+2\epsilon_\text{e}+\rho^3(\epsilon_\text{b}-\epsilon_\infty)]}{D}\\
 &p_{23}=\frac{3\epsilon_\text{e}(\epsilon_\text{b}-\epsilon_\infty)}{D}\\
 &p_{30}=-\frac{3\rho^3\epsilon_\infty}{D}
  \end{align*}
\begin{align*}
 &p_{31}=-\frac{(1-\rho^3)(\epsilon_\text{b}+2\epsilon_\infty)}{D}\\
 &p_{32}=\frac{2\rho^3(1-\rho^3)(\epsilon_\text{b}-\epsilon_\infty)}{D}\\
 &p_{33}=\frac{(\epsilon_\infty-\epsilon_\text{e})(\epsilon_\text{b}+2\epsilon_\infty)+\rho^3(\epsilon_\text{b}-\epsilon_\infty)(\epsilon_\text{e}+2\epsilon_\infty)}{D}.
\end{align*}

This gives us the system of Eqs.~\eqref{eq:p0pl}-\eqref{eq:p3pl} as written down in the main article.

\section{Exciting field and Emission Intensity}
We here give some indication on how the dynamics of the nanoshell in the lasing regime was obtained, as exposed in the ``Above threshold'' section of the main article.

Since we were interested in investigating situations of free lasing (no external drive), it seemed physical to use zero-field initial conditions, and zero external probe, and then leave the lasing instability to grow out of the numerical noise. This procedure, however, gives rise to prohibitively long computational times, and becomes especially inefficient when computing spectra including many frequency points. This is why as a numerical trick, we in fact applied a minute probe field $E_0$ acting like a ``seed'' and driving the initial steps of the instability faster. To produce the figures shown in Section~\ref{sc:abvth}, we chose to apply a field value $E_0=10^{-8} E_\text{sat}$.

To make sure nonetheless that the results we obtained were in the free lasing regime and that the presence of the small $E_0$ did not generate any forced oscillation regime, we verified that the final results did not depend on the value chosen for $E_0$. This is illustrated in the following Figure, where the emitted intensity $I_\text{em}(t)$ has been calculated in the same conditions as Figs.~\ref{fg:qabv} to \ref{fg:abv} of the main article, namely, at the frequency $\hbar\omega=2.811$~eV and with a gain level $G=1.01\,G_\text{th}$. Results are shown for $E_0= 10^{-10}$ to $10^{-7} E_\text{sat}$: it is seen that the obtained responses are indeed all exactly the same to within some time translation, corresponding to the onset time of the lasing instability.

\begin{figure}[ht]
\centering
\includegraphics[width=\textwidth]{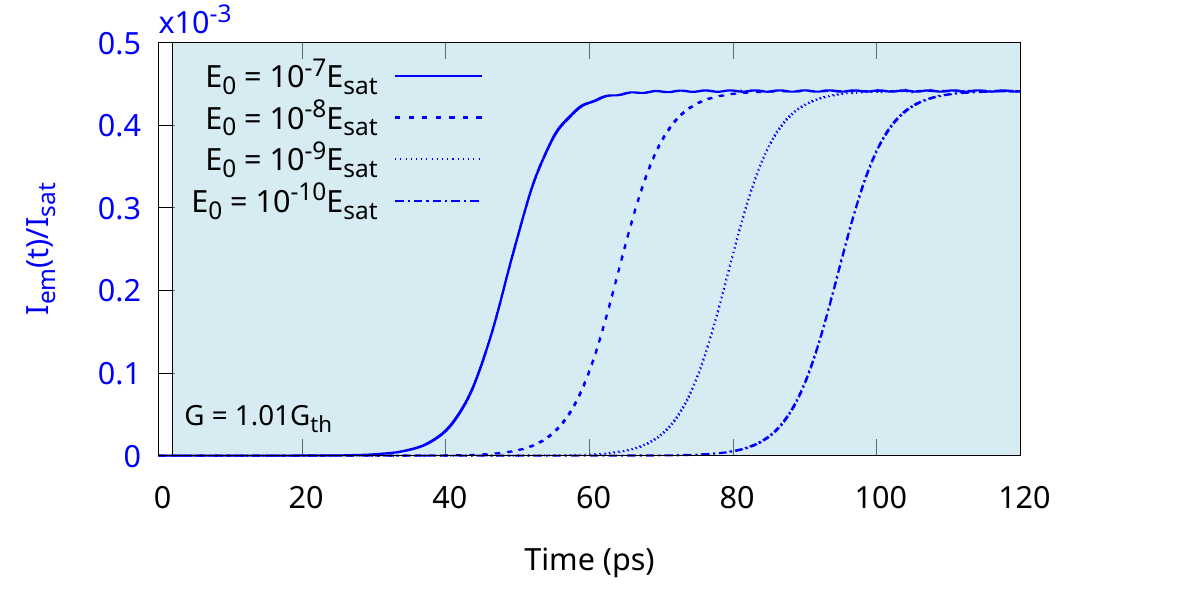}
\caption{Identical results (to within a time translation) were obtained for all values of the field $E_0$ used to accelerate the numerical onset of the lasing instability, confirming that the nanolaser is in a free lasing regime.}
\label{fg:rangeE0}
\end{figure}

\end{document}